%% file: main.tex
\documentclass[sigconf, nonacm]{acmart}

\usepackage{listings}
\usepackage{tikz}
\usetikzlibrary{arrows.meta,calc,positioning}
\definecolor{OneDarkBackground}{HTML}{282C34}
\definecolor{OneDarkForeground}{HTML}{ABB2BF}
\definecolor{OneDarkPurple}{HTML}{C678DD}
\definecolor{OneDarkBlue}{HTML}{61AFEF}
\definecolor{OneDarkGreen}{HTML}{98C379}
\definecolor{OneDarkGray}{HTML}{7F848E}
\definecolor{OneDarkRed}{HTML}{E06C75}
\definecolor{OneDarkYellow}{HTML}{E5C07B}
\definecolor{CodePurple}{HTML}{6A1B9A}
\definecolor{CodeBlue}{HTML}{0057A8}
\definecolor{CodeGreen}{HTML}{137333}
\definecolor{CodeGray}{HTML}{4D5963}
\definecolor{CodeRed}{HTML}{B3261E}
\definecolor{CodeBrown}{HTML}{8A5100}
\lstdefinestyle{onedarkpython}{
  language=Python,
  backgroundcolor=\color{white},
  basicstyle=\ttfamily\footnotesize\color{black!90},
  keywordstyle=\color{CodePurple}\bfseries,
  commentstyle=\color{CodeGray},
  stringstyle=\color{CodeGreen},
  emph=[1]{F,expand,reduce,reduce_aligned,render,ocr,metadata,assemble,forward,
    __init__},
  emphstyle=[1]\color{CodeBlue},
  emph=[2]{self},
  emphstyle=[2]\color{CodeBrown},
  emph=[3]{pdfs,pages,page_grains,ocr_results,ocr_results_list,stems,groups,
    ordered_pages,members},
  emphstyle=[3]\color{CodeRed},
  emph=[4]{PdfPipeline,Pipeline,RayModule,MinerUPdfToPages,MinerUVlmOcrPage,
    PdfMetadata,MinerUAssembleDoc},
  emphstyle=[4]\color{CodeBrown},
  showstringspaces=false,
  keepspaces=true,
  columns=fullflexible,
  breaklines=true,
  frame=single,
  framerule=0.4pt,
  rulecolor=\color{black!55},
  framesep=0pt,
  xleftmargin=0pt,
  xrightmargin=0pt,
  framexleftmargin=0pt,
  framexrightmargin=0pt,
  aboveskip=0pt,
  belowskip=0pt
}

\usepackage{xspace}
\usepackage{enumitem}
\usepackage{array}
\usepackage{fancyhdr}

\newcommand{\sys}{\textsc{RayOrch}\xspace}

\usepackage{quoting}
\quotingsetup{leftmargin=0.67cm, rightmargin=0.67cm}

\graphicspath{{figures/}}

\fancypagestyle{arxivlogos}{%
  \fancyhf{}%
  \fancyhead[C]{%
    \makebox[\textwidth][c]{%
      \includegraphics[height=0.32in]{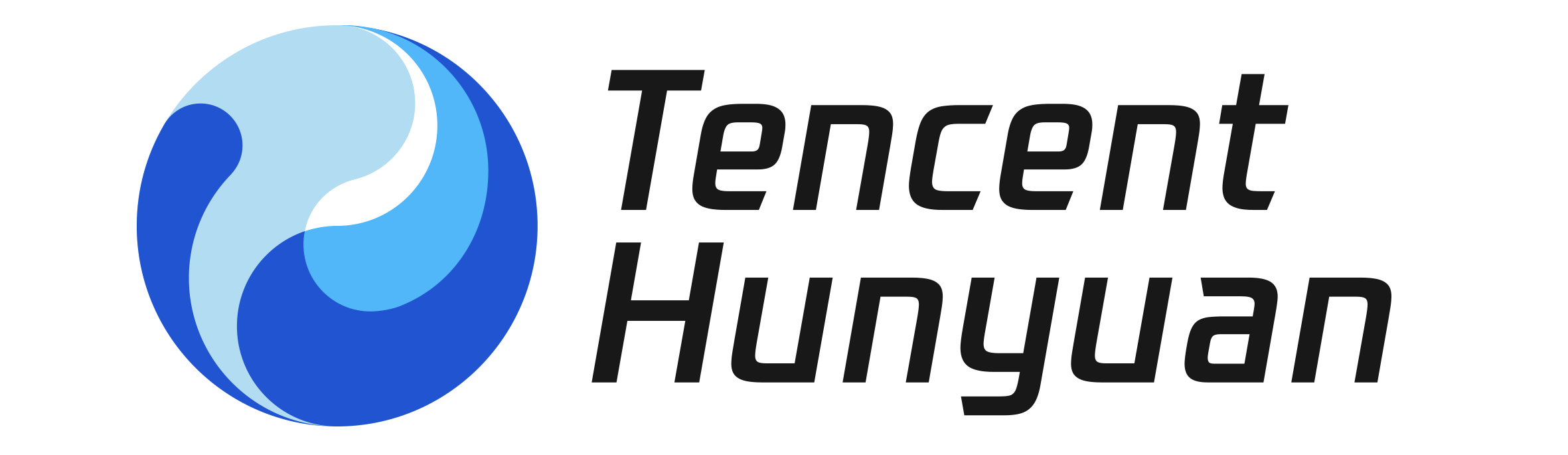}%
      \hspace{0.35in}%
      \includegraphics[height=0.30in]{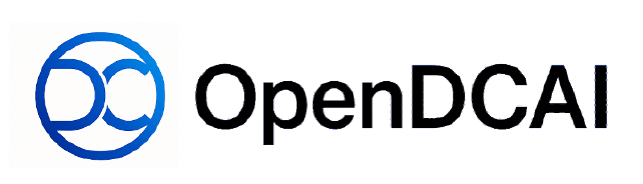}%
      \hspace{0.35in}%
      \includegraphics[height=0.34in]{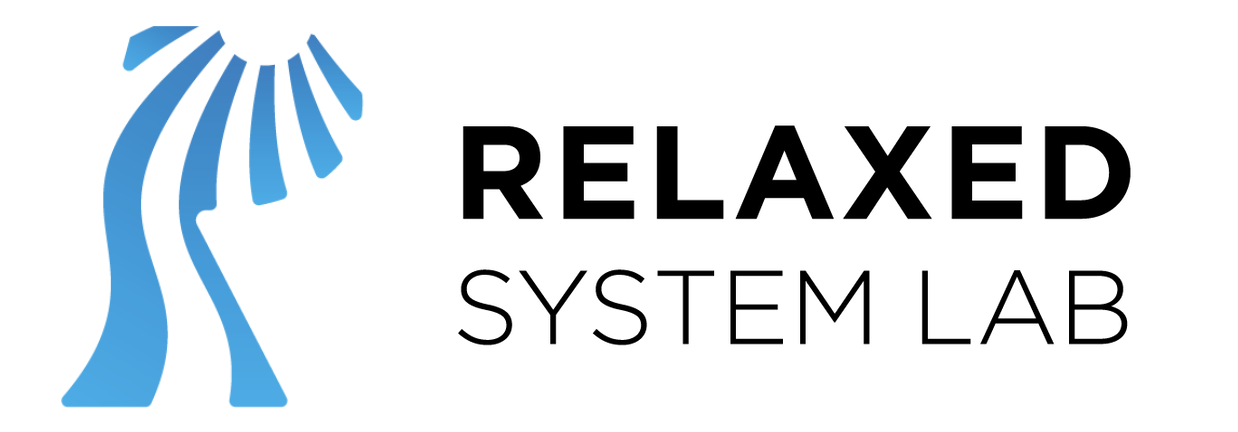}%
      \hspace{0.35in}%
      \includegraphics[height=0.24in]{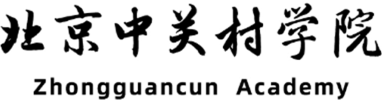}%
    }%
  }%
  \fancyfoot[C]{\thepage}%
}
\begin{document}

\title{\sys: Programming and Executing Lineage-Controlled Multi-Grain
Dataflows for Foundation-Model Data Preparation}

\author{Xiaochen Ma}
\author{Zimo Meng}
\author{Junzhu Liang}
\author{Youhe Jiang}
\author{Yue Cheng}
\author{Hao Liang}
\author{Bohan Zeng}
\author{Dengchun Li}
\author{Lu Ma}
\author{Zhengyang Zhao}
\author{Zhen Hao Wong}
\author{Runming He}
\author{Meiyi Qiang}
\author{Jiangtao Guan}
\author{Binhang Yuan}
\author{Wentao Zhang}
\renewcommand{\shortauthors}{Ma et al.}

\makeatletter
\gdef\addresses{%
  \@author{%
    Xiaochen Ma\textsuperscript{*},
    Zimo Meng\textsuperscript{*},
    Junzhu Liang\textsuperscript{*},
    Youhe Jiang\textsuperscript{*},
    Yue Cheng \\
    Hao Liang,
    Bohan Zeng,
    Dengchun Li,
    Lu Ma,
    Zhengyang Zhao,
    Zhen Hao Wong \\
    Runming He,
    Meiyi Qiang,
    Jiangtao Guan,
    Binhang Yuan\textsuperscript{$\dagger$},
    Wentao Zhang\textsuperscript{$\dagger$} \\[0.25em]
    {\small
      Peking University \quad HKUST \quad University of Cambridge \quad
      Tencent Hunyuan \quad Zhongguancun Academy} \\[-0.1em]
    {\footnotesize
      \textsuperscript{*}Equal contribution.\quad
      \textsuperscript{$\dagger$}Corresponding authors:
      \texttt{biyuan@ust.hk}, \texttt{wentao.zhang@pku.edu.cn}.}%
  }%
}
\makeatother


\begin{abstract}
\input{sections/0_abstract}
\end{abstract}

\maketitle
\thispagestyle{arxivlogos}

\input{sections/1_introduction}

\input{sections/2_related_work}

\input{sections/3_overview}

\input{sections/4_design}

\input{sections/5_implementation}
\input{sections/6_evaluation}

\input{sections/8_conclusion}

\bibliographystyle{ACM-Reference-Format}
\bibliography{main}

\end{document}

%% file: sections/0_abstract.tex
Preparing high-quality training data for foundation models requires scalable pipelines that transform large collections of heterogeneous documents or videos into structured training records. These pipelines could repeatedly change their unit of processing. At each expansion step, one input item (i.e., the
parent node in this graph) produces an ordered, input-dependent sequence of output items (viewed as its children in this data lineage graph), where the distribution of child counts is long-tailed across parents. On the other hand, the GPU running a particular data pipeline stage should batch children from different parents to maximize utilization, and the system must still return each result to its immediate parent, preserve child order, and determine when all required child results have become terminal.
Existing data-pipeline systems usually choose between two imperfect options.
Coarse-grained functions keep each document or video as one opaque job, hiding the pages, clips, or frames that could run in parallel. Flat-record functions expose these items individually, but force applications to remember each item's
parent and position, track when all items are finished, and globally regroup
the records to rebuild the original result.

To address these challenges, we present \sys, a programming model and distributed execution engine that maintains these parent-child relations throughout execution. A
program declares an ordered variable-cardinality parent-to-child expansion and
a matching child-to-parent gather that reconstructs each parent result. The compiler validates each
expansion-gather pair. At runtime, \sys records the structural lineage of
every expansion: its concrete child set, each child's immediate parent and
immutable ordinal, and each result's terminal state. Per-Call FIFO Ready Queues batch ready children across parents, while gathers use declared
membership and ordinals rather than batch boundaries or completion order.
A parent can therefore finalize its result and advance to the next stage as
soon as all required child results become terminal. When a Call reports a typed parent-scoped failure, \sys suppresses undispatched siblings of that parent for the Call while allowing
unrelated parents to continue.

To verify the design of \sys, we conduct comprehensive evaluations. On NVIDIA H20 GPUs, \sys achieves a 15.14$\times$ processing-time speedup
when scaling MinerU from 4 to 64 GPUs and a 7.82$\times$ speedup when scaling
the video pipeline from 8 to 64 GPUs. It reduces end-to-end time by 13.1\%
versus Ray Data and 29.0\% versus Daft on MinerU, and by 16.0\% versus Ray Data
on Docling. FIFO dispatch reduces ablation wall time from 634.1 to 579.3
seconds (8.6\%). In a controlled failure-injection experiment, \sys
prevents 6{,}241 of 23{,}514 nontrigger sibling computations from entering the
UDF and reduces wall time by 14.93\% on average relative to matched
failure-free runs while preserving all expected outputs for unaffected
parents. Code available at {\color{blue}\url{https://github.com/OpenDCAI/RayOrch}}.

%% file: sections/1_introduction.tex
\section{Introduction}
\label{sec:introduction}

\input{figures/teaser_lineage_pipeline_v2}

Foundation-model data preparation requires to efficiently process distributed dataflows at scale.
Pipelines for layout-rich documents and long videos combine
decoding, parsing, CPU transformations, GPU inference, and final assembly.
They also repeatedly change their unit of processing: a PDF may produce pages,
regions, and tables, while a video may produce clips, frames, and audio
segments
\cite{poznanski2025olmocr,li2024multimodal,laurencon2024building,bruce2024genie,nvidia2025cosmos}.
At corpus scale, an execution engine must expose the parallelism of these finer
units while returning every result to its immediate parent in the correct
order.  Otherwise, the system either limits child-level parallelism and
cross-parent batching or makes applications implement correctness-critical
grouping, ordering, and completion logic. In this paper, we tackle the concrete problem:

\vspace{0.35em}
\begin{quoting}
    \noindent\textit{How can a distributed dataflow system batch fine-grained work across parents while preserving the structure required for per-parent completion, ordered reconstruction, and failure containment when each parent produces an input-dependent, ordered sequence of children?}
\end{quoting}
\vspace{0.35em}

The intrinsic difficulty in this problem is that logical expansions and efficient physical
batches have different boundaries.  An \emph{ordered variable-cardinality
expansion} materializes, for each parent data unit, an input-dependent ordered
sequence of child data units; the distribution of child counts across parents
can be long-tailed, as Figure~\ref{fig:motivation}(a) illustrates.  A GPU stage
benefits from dispatch batches that mix ready computations from several
parents.  Yet the runtime must retain each child unit's immediate parent and
immutable ordinal, and a parent-scoped gather is resolvable only after expansion
membership is fixed and all required child outcomes are terminal.  Dispatch-batch
composition and completion order therefore cannot define expansion membership
or reconstruction order.  Moreover, a typed parent-scoped failure should stop
undispatched sibling computations for the same configured stage and parent
without impeding other parents.

Existing representations each capture only one side of this requirement.  A
coarse per-source function keeps the parent-child relation implicit but hides
child-level work from the dataflow engine, limiting its ability to schedule
children independently, provision child stages separately, and form
cross-parent model batches.  A flat representation exposes that parallelism
and batching, but turns structural facts into application data.  Distributed
engines such as Spark~\cite{zaharia2012rdd}, Dask~\cite{rocklin2015dask}, and
Ray~\cite{moritz2018ray} provide distributed collections and task graphs.
Ray Data~\cite{luan2025streaming} and Daft~\cite{eventual2026daft} add
pipelined execution, cardinality-changing operators, and cross-record batching.
When these systems express fan-out through flat collections, an application
can carry each child's immediate-parent key and immutable ordinal as fields.
The flat interfaces considered here do not interpret those fields structurally.
They leave concrete membership and terminal child outcomes as application data
rather than state that governs downstream readiness or gather resolution.
Applications can maintain counts or group and sort results, but must then define
parent completion and failure semantics themselves.  In our evaluated Ray Data and
Daft pipelines, this reconstruction also introduces a post-child-stage shuffle
or regrouping barrier.  The missing contract is a declared relation that
remains available to scheduling, ordered gathering, and failure containment
after children enter physical batches, as Figure~\ref{fig:motivation}(b)
illustrates.

Our key insight is to retain \emph{structural lineage state} throughout an
execution while allowing physical batches to remain transient.  This state
records each materialized expansion's concrete ordered child set, every child's
immediate parent and immutable zero-based ordinal, and the terminal outcomes of
the expansion and its child results.  The program declares the expansion and
gather relations, which the compiler validates.  The runtime materializes and
maintains the corresponding structural lineage state.  Here structural lineage
denotes control state for online scheduling and completion, not a post hoc
provenance record or Ray's task-recomputation lineage.  We realize this insight in
\sys, a programming model and distributed execution engine for
\emph{multi-grain dataflows} over finite, acyclic Domain trees.  A Domain names
a compile-time logical level whose runtime instances are Entities; a Call is
one configured use of a user Function; and applying a Call to one Entity forms
a \emph{Grain}, the unit of scheduling, retry, and commit.  A multi-grain
dataflow connects multiple Domains through paired ordered variable-cardinality
expansions and parent-scoped gathers.

A \sys program uses \texttt{F.expand} to declare a parent-to-child Domain
relation and materialize its ordered child Entities at runtime, and
\texttt{F.reduce} as the matching parent-scoped ordered gather.  The compiler
checks Domain compatibility and matches each pair.  Within each active source
microbatch, normal per-Call FIFO ready queues hold ready Grains in enqueue
order; the configured reservation policy forms physical batches, which may mix
parents.  Recovery uses separate priority queues.  Immutable ordinals govern
each gather; queue and completion order do not.  The runtime accepts a report
only for the matching in-flight Grain and generation.  An accepted report seals
that Grain and atomically publishes its terminal output facts; this generation
fence rejects stale or duplicate reports after retry or actor replacement.  A
typed \texttt{GroupFailure} installs a barrier for one Call and
immediate parent: ready sibling Grains are suppressed before dispatch, reports
from in-flight siblings are rejected, and unrelated parents remain live.
\sys supports ordered gathering within each source microbatch; it
does not support general joins, windows that span microbatches, or feedback.
Figure~\ref{fig:motivation}(c) summarizes this separation of structure from
execution.

We implement \sys as a Python layer on Ray and evaluate it on MinerU and
Docling document pipelines and a Qwen2.5-VL-7B video pipeline.  On NVIDIA H20
GPUs, \sys reduces end-to-end time by 13.1\% versus Ray Data and 29.0\%
versus Daft on MinerU, and by 16.0\% versus Ray Data on Docling.  Under strong
scaling, its processing-time speedup reaches 15.14$\times$ from 4 to 64 GPUs
on MinerU and 7.82$\times$ from 8 to 64 GPUs on video.  FIFO dispatch reduces
ablation wall time from 634.1 to 579.3 seconds (8.6\%).  In a controlled
failure-injection experiment, typed failure containment prevents 6{,}241 of
23{,}514 nontrigger sibling computations from entering the UDF and reduces wall
time by 14.93\% on average relative to matched failure-free runs, while
preserving every expected output for unaffected parents.
We summarize the following key contributions:

\begin{itemize}[leftmargin=*,nosep]
  \item We formulate a structural model for finite, acyclic hierarchical
  dataflows, which makes expansions and parent-scoped gathers explicit and
  statically checkable.
  \item We develop an engine that maintains lineage state, batches ready Grains across parents, and resolves parents independently of materialized membership and immutable ordinals.
  \item We specify generation-fenced per-Grain commit and typed failure
  containment scoped to one Call and immediate parent.
  \item We demonstrate end-to-end gains on MinerU and Docling, near-linear
  strong scaling through 64 GPUs, and measurable benefits from FIFO dispatch
  and typed failure containment.
\end{itemize}

%% file: figures/teaser_lineage_pipeline_v2.tex
\begin{figure*}[t]
\centering
\begingroup
\definecolor{RayTeaserA}{HTML}{4C78A8}
\definecolor{RayTeaserB}{HTML}{F28E2B}
\definecolor{RayTeaserC}{HTML}{2E9B61}
\definecolor{RayTeaserLineage}{HTML}{59A14F}
\definecolor{RayTeaserExec}{HTML}{7A6F9B}
\definecolor{RayTeaserBad}{HTML}{D9534F}
\resizebox{0.9\textwidth}{!}{%
\begin{tikzpicture}[
  x=1cm,
  y=1cm,
  font=\sffamily\scriptsize,
  flow/.style={-{Latex[length=1.8mm,width=1.2mm]},draw=black!62,
    line width=0.55pt},
  guide/.style={draw=black!38,dashed,line width=0.5pt},
  panel/.style={draw=black!22,fill=black!1,rounded corners=3pt,
    line width=0.55pt},
  sourceA/.style={draw=RayTeaserA,fill=RayTeaserA!9,rounded corners=2pt,
    minimum width=12mm,minimum height=6mm,inner sep=1.5pt,font=\bfseries},
  sourceB/.style={draw=RayTeaserB,fill=RayTeaserB!9,rounded corners=2pt,
    minimum width=12mm,minimum height=6mm,inner sep=1.5pt,font=\bfseries},
  sourceC/.style={draw=RayTeaserC,fill=RayTeaserC!9,rounded corners=2pt,
    minimum width=12mm,minimum height=6mm,inner sep=1.5pt,font=\bfseries},
  grainA/.style={draw=RayTeaserA,fill=RayTeaserA!13,rounded corners=1.5pt,
    minimum width=5.3mm,minimum height=5.2mm,inner sep=0.7pt},
  grainB/.style={draw=RayTeaserB,fill=RayTeaserB!14,rounded corners=1.5pt,
    minimum width=5.3mm,minimum height=5.2mm,inner sep=0.7pt},
  grainC/.style={draw=RayTeaserC,fill=RayTeaserC!13,rounded corners=1.5pt,
    minimum width=5.3mm,minimum height=5.2mm,inner sep=0.7pt},
  batch/.style={draw=RayTeaserExec!72!black,fill=RayTeaserExec!5,
    dashed,rounded corners=2pt,minimum width=2.60cm,minimum height=7.2mm},
  fifo/.style={draw=black!50,fill=white,rounded corners=2pt,
    minimum width=5.45cm,minimum height=7.5mm},
  status/.style={draw=black!26,fill=white,rounded corners=2pt,
    minimum width=1.25cm,minimum height=5.7mm,inner sep=1.1pt,font=\bfseries},
  lineagekey/.style={draw=RayTeaserLineage!68!black,
    fill=RayTeaserLineage!8,rounded corners=1.5pt,
    minimum width=1.30cm,minimum height=4.8mm,align=center,
    inner xsep=2.2pt,inner ysep=0.8pt,
    font=\sffamily\scriptsize\bfseries,text=RayTeaserLineage!62!black},
  nextA/.style={draw=RayTeaserA,fill=RayTeaserA!9,rounded corners=2pt,
    minimum height=5.3mm,minimum width=1.08cm,inner sep=0.8pt,font=\bfseries},
  waitB/.style={draw=RayTeaserB,fill=RayTeaserB!9,rounded corners=2pt,
    minimum height=5.3mm,minimum width=1.08cm,inner sep=0.8pt,font=\bfseries},
  nextC/.style={draw=RayTeaserC,fill=RayTeaserC!9,rounded corners=2pt,
    minimum height=5.3mm,minimum width=1.08cm,inner sep=0.8pt,font=\bfseries},
  title/.style={anchor=west,font=\sffamily\small\bfseries},
  note/.style={font=\sffamily\scriptsize,text=black!67}
]

\draw[panel] (0,0.88) rectangle (6.00,5.82);
\draw[panel] (6.20,0.88) rectangle (12.40,5.82);
\draw[panel] (12.60,0.88) rectangle (19.40,5.82);
\node[title] at (0.20,5.50) {(a) PDF/video fan-out is long-tailed};
\node[title] at (6.40,5.50) {(b) Flat rebatching (Ray Data, Daft)};
\node[title] at (12.80,5.50) {(c) Lineage + FIFO continuation (ours)};

\node[sourceA] (srcA) at (0.78,4.00) {PDF $A$};
\node[sourceB] (srcB) at (0.78,3.00) {PDF $B$};
\node[sourceC] (srcC) at (0.78,2.00) {VIDEO $C$};

\node[grainA] (a0) at (2.03,4.00) {$A_0$};
\node[grainA] at (2.69,4.00) {$A_1$};
\draw[flow] (srcA.east) -- (a0.west);
\node[note,anchor=east,font=\bfseries,text=RayTeaserA!82!black]
  at (5.72,4.00) {2 pages};

\node[grainB] (b0) at (2.03,3.00) {$B_0$};
\node[grainB] at (2.69,3.00) {$B_1$};
\node[grainB] at (3.35,3.00) {$B_2$};
\node[note] at (3.70,3.00) {$\cdots$};
\node[grainB] at (4.06,3.00) {$B_{47}$};
\draw[flow] (srcB.east) -- (b0.west);
\node[note,anchor=east,font=\bfseries,text=RayTeaserB!84!black]
  at (5.82,3.00) {48 pages};

\node[grainC] (c0) at (2.03,2.00) {$C_0$};
\node[grainC] at (2.69,2.00) {$C_1$};
\node[grainC] at (3.35,2.00) {$C_2$};
\node[note] at (3.70,2.00) {$\cdots$};
\node[grainC] at (4.06,2.00) {$C_{119}$};
\draw[flow] (srcC.east) -- (c0.west);
\node[note,anchor=east,font=\bfseries,text=RayTeaserC!82!black]
  at (5.82,2.00) {120 clips};

\node[note,align=center,font=\bfseries] at (3.00,1.42)
  {pages $\rightarrow$ regions; clips $\rightarrow$ frames/audio};
\node[note,align=center] at (3.00,1.10)
  {input-dependent counts: most short, a few huge};

\node[note,anchor=west] at (6.42,4.76)
  {mixed batches are already available};
\node[batch] at (7.92,3.90) {};
\node[grainA] at (7.01,3.90) {$A_0$};
\node[grainB] at (7.62,3.90) {$B_0$};
\node[grainC] at (8.23,3.90) {$C_0$};
\node[grainA] at (8.84,3.90) {$A_1$};
\node[batch] at (10.72,3.90) {};
\node[grainB] at (9.81,3.90) {$B_1$};
\node[grainC] at (10.42,3.90) {$C_1$};
\node[grainB] at (11.03,3.90) {$B_2$};
\node[grainC] at (11.64,3.90) {$C_2$};

\node[status,text=RayTeaserA!84!black] (flatA) at (7.35,3.08) {$A:2/2$};
\node[status,text=RayTeaserC!82!black] (flatC) at (9.30,3.08) {$C:3/3$};
\node[status,text=RayTeaserB!86!black] (flatB) at (11.25,3.08) {$B:3/5$};

\node[draw=black!48,fill=white,rounded corners=2pt,
  minimum width=4.85cm,minimum height=7.2mm,align=center,font=\bfseries]
  (globalreduce) at (9.30,2.17) {GLOBAL GROUPBY $+$ SORT $+$ REDUCE};
\draw[flow] (flatA.south) -- (globalreduce.north west);
\draw[flow] (flatC.south) -- (globalreduce.north);
\draw[flow] (flatB.south) -- (globalreduce.north east);
\node[note] at (9.30,1.69)
  {reconstruct membership, order, and completion};
\node[draw=RayTeaserBad!72!black,fill=RayTeaserBad!6,rounded corners=2pt,
  minimum width=5.15cm,minimum height=5.8mm,align=center,font=\bfseries,
  text=RayTeaserBad!78!black] at (9.30,1.18)
  {$A$ and $C$ wait for $B$; downstream actors idle};

\node[note,anchor=west,font=\sffamily\tiny\bfseries,
  text=RayTeaserLineage!64!black] at (12.72,4.87) {lineage};
\node[lineagekey] at (14.50,4.87) {parent};
\node[lineagekey] at (15.88,4.87) {ordinal};
\node[lineagekey] at (17.26,4.87) {membership};
\node[lineagekey] at (18.64,4.87) {completion};

\node[note,font=\bfseries,fill=white,inner sep=0.7pt]
  at (16.00,4.32) {per-Call FIFO Ready Queue};
\node[fifo] (fifo) at (16.00,3.78) {};
\foreach \x/\style/\text in {
  13.70/grainA/$A_0$,14.34/grainB/$B_0$,14.98/grainC/$C_0$,
  15.62/grainA/$A_1$,16.26/grainB/$B_1$,16.90/grainC/$C_1$,
  17.54/grainB/$B_2$,18.18/grainC/$C_2$}
  {\node[\style] at (\x,3.78) {\text};}
\node[note,anchor=east,font=\sffamily\tiny,fill=white,inner sep=0.5pt]
  at (13.08,3.78) {append};
\draw[flow] (13.14,3.78) -- (fifo.west);
\node[note,anchor=west,font=\sffamily\tiny,fill=white,inner sep=0.5pt]
  at (18.78,3.78) {reserve};
\draw[flow] (fifo.east) -- (18.76,3.78);

\node[batch] at (14.58,2.62) {};
\node[grainA] at (13.67,2.62) {$A_0$};
\node[grainB] at (14.28,2.62) {$B_0$};
\node[grainC] at (14.89,2.62) {$C_0$};
\node[grainA] at (15.50,2.62) {$A_1$};
\node[batch] at (17.42,2.62) {};
\node[grainB] at (16.51,2.62) {$B_1$};
\node[grainC] at (17.12,2.62) {$C_1$};
\node[grainB] at (17.73,2.62) {$B_2$};
\node[grainC] at (18.34,2.62) {$C_2$};
\node[note,above] at (14.58,3.06) {dense batch 1};
\node[note,above] at (17.42,3.06) {dense batch 2};
\draw[flow] (15.44,3.40) -- (14.82,3.03);
\draw[flow] (16.56,3.40) -- (17.18,3.03);

\node[font=\sffamily\tiny\bfseries,text=RayTeaserA!84!black,anchor=east]
  (doneA) at (13.55,1.72) {$A:2/2$};
\node[nextA,font=\sffamily\tiny\bfseries] (nextA) at (14.25,1.72) {Next($A$)};
\draw[flow,draw=RayTeaserA] (doneA.east) -- (nextA.west);
\node[font=\sffamily\tiny\bfseries,text=RayTeaserB!86!black,anchor=east]
  (doneB) at (15.75,1.72) {$B:3/5$};
\node[waitB,font=\sffamily\tiny\bfseries] (waitB) at (16.45,1.72) {waiting};
\draw[flow,draw=RayTeaserB] (doneB.east) -- (waitB.west);
\node[font=\sffamily\tiny\bfseries,text=RayTeaserC!82!black,anchor=east]
  (doneC) at (17.95,1.72) {$C:3/3$};
\node[nextC,font=\sffamily\tiny\bfseries] (nextC) at (18.65,1.72) {Next($C$)};
\draw[flow,draw=RayTeaserC] (doneC.east) -- (nextC.west);
\node[note,anchor=west] at (12.88,1.16)
  {completed parents continue immediately};

\draw[RayTeaserLineage!72!black,line width=0.7pt]
  (0.20,0.38) -- (4.22,0.38);
\node[font=\bfseries,text=RayTeaserLineage!58!black] at (9.70,0.38)
  {Persistent lineage makes cross-parent batching safe and reconstruction local.};
\draw[RayTeaserLineage!72!black,line width=0.7pt]
  (15.18,0.38) -- (19.20,0.38);

\end{tikzpicture}%
}
\endgroup

\vspace{-1.5em}
\caption{Why long-tailed multi-grain pipelines need lineage control.  (a) Sources
fan out dynamically.  (b) Flat rebatching fills batches but global
reconstruction couples parents.  (c) Runtime-owned lineage and per-Call FIFO
Ready Queues let complete parents continue independently.}
\Description{A three-panel teaser.  Panel a shows two PDFs and one video
producing 2 pages, 48 pages, and 120 clips, illustrating long-tailed changes in
data units. Panel b shows dense mixed batches followed by a global group-by,
sort, and reduce barrier at which completed parents A and C wait for incomplete
parent B.  Panel c shows a fine-grained lineage band above a per-Call FIFO Ready
Queue with dense mixed batches; four compact lineage
attributes identify the runtime-owned structure.  A and C continue to their next
stages while B remains incomplete.}
\label{fig:motivation}
\vspace{-1.5em}
\end{figure*}

%% file: sections/2_related_work.tex
\input{tables/system_positioning}

\vspace{-1.0em}
\section{Background and Related Work}
\label{sec:related-work}

\subsection{Multimodal Data Preparation}

Document-parsing systems transform PDFs or page images into structured text
and layout representations for downstream training or inference.  MinerU
combines specialized extraction models with preprocessing and postprocessing
rules~\cite{wang2024mineru}, whereas MinerU2.5 separates global layout analysis
from native-resolution recognition of selected regions~\cite{niu2026mineru2}.
Docling combines layout and table analysis in a conversion
toolkit~\cite{auer2024docling}, and SmolDocling provides a
compact end-to-end vision-language model for document
conversion~\cite{nassar2025smoldocling}.  Dolphin first generates layout
elements in reading order and then parses the corresponding elements in
parallel~\cite{feng2025dolphin}.  Although these systems differ in model
architecture, their distributed implementations repeatedly change the unit of
data: a document yields an input-dependent number of pages, and a page may
yield an input-dependent number of regions or other elements.  

Video preparation creates a variable number of units per source.  Wan's
data pipeline includes filtering, scoring, and dense
captioning~\cite{wan2025video}; Cosmos explicitly splits
each video into shots before filtering, annotation, deduplication, and
sharding~\cite{nvidia2025cosmos}.  A root video therefore produces an
input-dependent sequence of clips, while individual model operators may
process sampled frames or clip-level representations.  Applications that
assemble or audit source-level results must retain the immediate-parent
association and sequence position of each derived unit even when accelerator
batches mix units from different videos.  In contrast, large text-curation
pipelines such as those for Llama~3 and FineWeb emphasize staged filtering and
deduplication over predominantly flat document
records~\cite{grattafiori2024llama3,penedo2024fineweb}.  

\vspace{-1.0em}
\subsection{Pipeline Execution Systems}

General-purpose frameworks expose different layers of distributed execution.
Spark provides fault-tolerant distributed collections~\cite{zaharia2012rdd},
Dask represents computations as task graphs~\cite{rocklin2015dask}, and Ray
unifies distributed tasks and actors~\cite{moritz2018ray}.  At the dataset
layer, Ray Data implements a streaming-batch execution model for heterogeneous
pipelines~\cite{luan2025streaming}, while Daft provides a distributed
DataFrame interface for multimodal data and
cardinality-changing operators~\cite{eventual2026daft}.  These systems can
pipeline operators and expose parallel work after fan-out; the relevant
distinction is how the resulting parent-child relation is represented during
execution.

\sys instead includes each ordered variable-cardinality expansion in its
programming and runtime contract.  The program declares the relation, the
compiler validates it, and the runtime maintains each materialized instance.
Its structural lineage state records the concrete Expansion membership, each
child Entity's immediate parent and immutable zero-based ordinal, and the
terminal outcomes of associated Items.  A configured Call applied to an
Entity forms a Grain; only ready Grains are dispatched, and physical batches
may mix Grains from different parents.  The same structural state determines
when \texttt{F.reduce}, a parent-scoped ordered gather, can emit a
parent-domain Item and which same-Call sibling Grains are affected by typed
same-parent failure containment.

AI-oriented data systems address complementary layers.  Data-Juicer provides
operator libraries and infrastructure for foundation-model data
curation~\cite{chen2024datajuicer, chen2025datajuicer}, while Mixtera provides a
data plane for declarative training-data selection and mixtures ~\cite{bother2026mixtera} and Modyn orchestrates data-centric
ML pipelines~\cite{bother2025modyn}. \texttt{tf.data} supplies an ML input
processing framework~\cite{murray2021tfdata};
FastFlow accelerates input pipelines through distributed
offloading~\cite{um2023fastflow};cedar optimizes unified input
pipelines~\cite{zhao2024cedar}; and Pecan selects transformation order and
execution placement~\cite{graur2024pecan}.  Trident adapts operator configurations, parallelism, and placement atop
Ray Data~\cite{pan2026trident}, extending a broader line of adaptive stream-processing
systems~\cite{floratou2017dhalion,lian2023conttune,an2026dflop}.

\vspace{-1.0em}
\subsection{Structure, Progress, and Provenance}

Research effort on nested data also preserves logical nesting over a flat physical
representation.  Smith et al. compile nested
collection programs into semantically equivalent shredded queries and add
skew-aware execution~\cite{smith2021nested}.  Their focus is relational query
processing over nested collections; \sys instead uses each materialized
Expansion as online state for scheduling black-box CPU or GPU Calls,
resolving a parent-scoped ordered gather, and containing typed failures.
Other dataflow systems track different notions of progress.  Dataflow uses
event-time windows, watermarks, and triggers for unbounded, out-of-order
streams~\cite{akidau2015dataflow}, whereas Naiad tracks progress through
logical timestamps in iterative and streaming
computations~\cite{murray2013naiad}.  Their progress coordinates are temporal
or iterative; they do not denote the concrete ordered child sequence of one
materialized parent Expansion.  Titian, in turn, records data-level
provenance in Spark for forward and backward tracing~\cite{interlandi2015titian}.
Prior stream-processing systems also maintain operator state to support
scale-out and fault tolerance~\cite{castro2013integrating}.

%% file: tables/system_positioning.tex
\begin{table*}[t!]
\caption{Ordered variable-cardinality expansion across distributed systems.}
\vspace{-1.0em}
\label{tab:system-positioning}
\small
\centering
\setlength{\tabcolsep}{1.6pt}
\renewcommand{\arraystretch}{1.12}
\begin{tabular}{@{}>{\raggedright\arraybackslash}m{0.070\textwidth}|
  >{\centering\arraybackslash}m{0.082\textwidth}|
  >{\centering\arraybackslash}m{0.078\textwidth}|
  >{\centering\arraybackslash}m{0.104\textwidth}|
  >{\centering\arraybackslash}m{0.170\textwidth}|
  >{\centering\arraybackslash}m{0.132\textwidth}|
  >{\centering\arraybackslash}m{0.097\textwidth}|
  >{\centering\arraybackslash}m{0.092\textwidth}|
  >{\centering\arraybackslash}m{0.097\textwidth}@{}}
\hline
\multicolumn{1}{@{}c|}{} &
\multicolumn{1}{c|}{\textbf{Execution}} &
\multicolumn{2}{c|}{\textbf{Programming}} &
\multicolumn{2}{c|}{\textbf{After fan-out}} &
\multicolumn{2}{c|}{\textbf{Resolution}} &
\multicolumn{1}{c@{}}{\textbf{Failure handling}} \\ \hline
\textbf{System} &
\shortstack{\textbf{Pipeline}\\\textbf{parallelism}} &
\shortstack{\textbf{Fan-out}\\\textbf{API}} &
\shortstack{\textbf{Expansion}\\\textbf{maintained by}} &
\shortstack{\textbf{Parent/order}\\\textbf{represented as}} &
\shortstack{\textbf{Cross-parent}\\\textbf{batch formation}} &
\shortstack{\textbf{Parent gather}\\\textbf{resolved by}} &
\shortstack{\textbf{Ordered}\\\textbf{gather}} &
\shortstack{\textbf{Containment}\\\textbf{scope}} \\ \hline
\textbf{Ray Data} & \ding{51} & \texttt{flat\_map} &
\emph{Application} & Application fields &
Blocks + row count & \emph{Application} &
Group + sort & \emph{Application-defined} \\ \hline
\textbf{Daft} & \ding{51} & \texttt{explode} &
\emph{Application} & Application fields &
Row rebatching & \emph{Application} &
Group + sort & \emph{Application-defined} \\ \hline
\textbf{\sys} & \ding{51} & \texttt{F.expand} &
\shortstack{\emph{Program}\\\emph{+ runtime}} &
Child Entity identity &
Per-Call Ready Queue & Expansion state &
\texttt{F.reduce} &
\shortstack{Grain or\\Call$\times$parent} \\ \hline
\multicolumn{9}{@{}p{\textwidth}@{}}{\scriptsize
\textit{Legend.} \ding{51}: system-provided;
\emph{Application}: application-maintained; \emph{Program/runtime}:
program-declared, compiler-validated, and runtime-maintained.  Ray Data forms
physical blocks by row count, Daft re-batches rows by size, and \sys draws
dependency-ready Grains from a per-Call FIFO Ready Queue that may span
parents.} \\ \hline
\end{tabular}
\vspace{-1.5em}
\end{table*}

%% file: sections/3_overview.tex
\vspace{-1.0em}
\section{System Overview}
\label{sec:overview}

\input{figures/pdf_workflow_control_operators_draft}


RayOrch exposes each pipeline through two complementary views.  The logical view
describes the data units and structural transformations declared by the program,
while the execution view schedules their computations on available workers.
The logical structure remains stable as batching, retries, and worker placement change.
Figure~\ref{fig:program-control-execution} connects these views: its upper
plane shows the declared program structure, and its lower plane shows how the
same structure is executed.
We use a MinerU-style document pipeline as a running example
\cite{wang2024mineru}.  Rendering converts a PDF into an ordered,
variable-length list of pages; MinerU's 1.2B-parameter vision-language OCR
model consumes one page image at a time and emits a structured OCR result; and a
final materialization stage collects the page-level results into an ordered
Markdown document and its corresponding images.  A video pipeline follows the same
pattern, replacing pages with clips, frames, or audio segments.  The example
captures the central challenge: the pipeline gains parallelism by changing its
data unit, but the final result must still preserve ownership, order, and
completeness.

\vspace{-1.0em}
\subsection{Logical Dataflow}
\label{sec:overview-logical}

RayOrch represents this structure as a logical dataflow over hierarchical data
units.  A \emph{Domain} names one level of the hierarchy, such as PDFs or
Pages, and an \emph{Entity} is one unit at that level, such as document $A$ or
page $A_0$.  A \emph{Function} is reusable user code, while a \emph{Call} is
one configured use of that function.  Applying a Call to one Entity forms a
\emph{Grain}, the unit of computation.  A declared \emph{Port} carries logical
values; the value for one Port and one Entity is an \emph{Item}.  Table~\ref{tab:logical-objects}
summarizes these terms using the page-OCR example.

\input{tables/logical_objects}

Figure~\ref{fig:program-control-execution} presents both views: panel (a) is
the concise, Torch-like user interface, while panel (b)'s upper plane shows
the same structure independently of execution.

In the running example, rendering produces one Item containing a page list for
each PDF.  The structural operator \texttt{F.expand} changes the unit of
computation by exposing the pages in that list as ordered Page Entities.  The
OCR Call then creates one Grain for each page.  After the page results are
produced, \texttt{F.reduce} reconstructs the document-level Item in the
declared order.

The important point is that the expansion relation is declared by the
program.  It determines which child Entities belong to a parent and how they
are reconstructed; reduction does not rediscover these groups from record
fields or application-managed identifiers.  Physical execution only supplies
outcomes for the declared children.

This programming model is what we call a \emph{multi-grain dataflow}: a
pipeline in which logical data units change across stages while their
structural relation remains explicit. Users can therefore express page-level
or frame-level parallelism without manually maintaining child counts, grouping
state, or sorting logic.
Together, these logical objects and structural relations define the meaning of
the program. Their realization through lineage state, queues, batching, and
recovery is described in Section~\ref{sec:design}.

\vspace{-1.0em}
\subsection{Physical Execution}
\label{sec:overview-execution}

The lower plane of Figure~\ref{fig:program-control-execution}(b) shows the
physical realization of the Calls declared above.  Each configured Call
becomes an Operator with a per-Call FIFO Ready Queue and a pool of CPU or GPU
workers.  Ready Grains may be packed into transient physical batches,
including batches that mix parents; this is an execution choice rather than a
logical grouping.

The runtime retains each Grain's logical identity while executing it, so a
mixed-batch result can still be routed to its parent.  Once a
parent's declared children are complete, its next-stage work can start without
waiting for unrelated parents.  Plan validation, queues, batching, retries,
and commit rules are developed in Section~\ref{sec:design}.

%% file: figures/pdf_workflow_control_operators_draft.tex
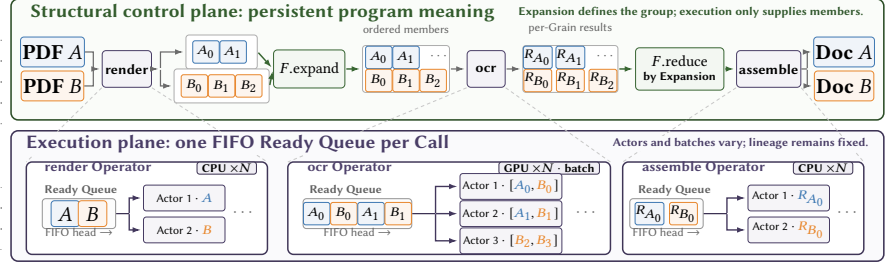
\begin{figure*}[t]
\centering
\begingroup
\definecolor{RayCtrl}{HTML}{4F8F47}
\definecolor{RayExec}{HTML}{66558D}
\definecolor{RayA}{HTML}{3E78B2}
\definecolor{RayB}{HTML}{E8872D}

\begin{minipage}[t]{0.34\textwidth}
\vspace{0pt}
{\sffamily\small\bfseries (a) RayOrch program}\par
\vspace{0.25em}
\begin{tikzpicture}
\node[
  draw=black!52,
  dash pattern=on 2.2pt off 1.5pt,
  line width=0.4pt,
  inner sep=0pt,
  outer sep=0pt,
  text width=\dimexpr\linewidth-0.8pt\relax,
  anchor=north west
] {
\begin{lstlisting}[
    style=onedarkpython,
    basicstyle=\ttfamily\fontsize{5.8}{6.9}\selectfont\color{black!90},
    frame=none,
    linewidth=\linewidth,
    xleftmargin=0pt,
    xrightmargin=0pt,
    breaklines=false,
    lineskip=0pt
  ]
class PdfPipeline(Pipeline):
  def __init__(self):
    self.render = RayModule(MinerUPdfToPages)
    self.ocr = RayModule(MinerUVlmOcrPage)
    self.assemble = RayModule(MinerUAssembleDoc)

  def forward(self, pdfs):
    pages = self.render(pdfs)
    page_grains = F.expand(pages)
    ocr_results = self.ocr(page_grains)
    groups = F.reduce(ocr_results)
    return self.assemble(groups, pages)
\end{lstlisting}};
\end{tikzpicture}
\end{minipage}%
\hfill
\begin{minipage}[t]{0.65\textwidth}
\vspace{0pt}
\centering
\resizebox{\linewidth}{!}{%
\begin{tikzpicture}[
  x=1cm,y=1cm,
  font=\sffamily\scriptsize,
  >={Latex[length=1.25mm,width=0.85mm]},
  flow/.style={->,draw=black!58,line width=0.52pt,
    shorten <=0.4pt,shorten >=0.8pt},
  guide/.style={draw=black!42,densely dotted,line width=0.55pt},
  ctrlflow/.style={->,draw=RayCtrl!72!black,line width=0.62pt,
    shorten <=0.4pt,shorten >=0.8pt},
  junction/.style={draw=RayExec!58!black,line width=0.52pt},
  branch/.style={->,draw=RayExec!65!black,line width=0.58pt,
    shorten <=0.2pt,shorten >=0.8pt},
  lowering/.style={draw=black!22,dash pattern=on 1.5pt off 1.25pt,
    line width=0.38pt},
  plane/.style={rounded corners=3pt,line width=0.65pt},
  title/.style={anchor=west,font=\sffamily\small\bfseries},
  note/.style={font=\sffamily\tiny,text=black!62},
  compute/.style={draw=RayExec!72!black,fill=RayExec!7,
    rounded corners=1.6pt,minimum width=9mm,minimum height=5.4mm,
    align=center,font=\sffamily\scriptsize\bfseries},
  structural/.style={draw=RayCtrl!70!black,fill=RayCtrl!7,
    rounded corners=1.6pt,minimum width=12mm,minimum height=5.6mm,
    align=center,font=\sffamily\scriptsize\bfseries},
  parentA/.style={draw=RayA,fill=RayA!8,rounded corners=1.4pt,
    minimum width=6.6mm,minimum height=4.4mm,inner sep=0.8pt,font=\bfseries},
  parentB/.style={draw=RayB,fill=RayB!8,rounded corners=1.4pt,
    minimum width=6.6mm,minimum height=4.4mm,inner sep=0.8pt,font=\bfseries},
  grainA/.style={draw=RayA,fill=RayA!8,rounded corners=1.0pt,
    minimum width=3.9mm,minimum height=3.3mm,inner sep=0.25pt,font=\sffamily\tiny},
  grainB/.style={draw=RayB,fill=RayB!8,rounded corners=1.0pt,
    minimum width=3.9mm,minimum height=3.3mm,inner sep=0.25pt,font=\sffamily\tiny},
  operator/.style={draw=RayExec!62!black,fill=white,rounded corners=2.4pt,
    line width=0.62pt,minimum height=13.6mm},
  fifo/.style={draw=black!42,fill=black!1,rounded corners=1.4pt,
    minimum height=4.4mm},
  actor/.style={draw=RayExec!70!black,fill=RayExec!6,rounded corners=1.2pt,
    minimum width=12.0mm,minimum height=3.8mm,align=center,
    inner sep=0.7pt,font=\sffamily\tiny},
  badge/.style={draw=RayExec!55!black,fill=RayExec!10,rounded corners=1.1pt,
    inner xsep=2pt,inner ysep=0.7pt,font=\sffamily\tiny\bfseries}
]

\node[title] at (0,4.12) {(b) Persistent structure and per-Call execution};

\draw[plane,draw=RayCtrl!54!black,fill=RayCtrl!3]
  (0,2.08) rectangle (12.85,3.86);
\node[title,text=RayCtrl!64!black] at (0.18,3.66)
  {Structural control plane: persistent program meaning};
\node[note,anchor=east,text=RayCtrl!60!black,font=\sffamily\tiny\bfseries]
  at (12.66,3.66) {Expansion defines the group; execution only supplies members.};

\draw[plane,draw=RayExec!60!black,fill=RayExec!3]
  (0,0) rectangle (12.85,1.92);
\node[parentA] (pdfa) at (0.61,3.08) {PDF $A$};
\node[parentB] (pdfb) at (0.61,2.58) {PDF $B$};
\node[compute,minimum width=7.2mm,inner xsep=1.2pt] (render)
  at (1.705,2.83) {render};

\node[draw=black!32,fill=white,rounded corners=1.3pt,
  minimum width=1.05cm,minimum height=5.0mm] (listA) at (3.10,3.08) {};
\node[grainA] at (2.87,3.08) {$A_0$};
\node[grainA] at (3.27,3.08) {$A_1$};
\node[draw=black!32,fill=white,rounded corners=1.3pt,
  minimum width=1.40cm,minimum height=5.0mm] (listB) at (3.10,2.58) {};
\node[grainB] at (2.70,2.58) {$B_0$};
\node[grainB] at (3.10,2.58) {$B_1$};
\node[grainB] at (3.50,2.58) {$B_2$};

\node[structural,minimum width=10.5mm,inner xsep=1.4pt] (expand)
  at (4.39,2.83) {$F.\mathrm{expand}$};
\node[draw=black!32,fill=white,rounded corners=1.3pt,
  minimum width=1.28cm,minimum height=6.9mm] (members) at (5.815,2.83) {};
\node[grainA] at (5.415,3.00) {$A_0$};
\node[grainA] at (5.815,3.00) {$A_1$};
\node[note,anchor=west] at (6.055,3.00) {$\cdots$};
\node[grainB] at (5.415,2.66) {$B_0$};
\node[grainB] at (5.815,2.66) {$B_1$};
\node[grainB] at (6.215,2.66) {$B_2$};
\node[note,anchor=south] at (5.815,3.24) {ordered members};

\node[compute,minimum width=5.6mm,inner xsep=0.8pt] (ocr)
  at (6.995,2.83) {ocr};
\node[draw=black!32,fill=white,rounded corners=1.3pt,
  minimum width=1.40cm,minimum height=6.9mm] (results) at (8.235,2.83) {};
\node[grainA] at (7.755,3.00) {$R_{A_0}$};
\node[grainA] at (8.235,3.00) {$R_{A_1}$};
\node[note,anchor=west] at (8.505,3.00) {$\cdots$};
\node[grainB] at (7.755,2.66) {$R_{B_0}$};
\node[grainB] at (8.235,2.66) {$R_{B_1}$};
\node[grainB] at (8.715,2.66) {$R_{B_2}$};
\node[note,anchor=south] at (8.235,3.24) {per-Grain results};

\node[structural,minimum width=13mm,inner xsep=1.2pt] (reduce) at (9.845,2.83)
  {$F.\mathrm{reduce}$\\[-1pt]\tiny by Expansion};
\node[compute,minimum width=7.8mm,inner xsep=1.0pt] (assemble)
  at (11.155,2.83) {assemble};
\node[parentA] (doca) at (12.265,3.08) {Doc $A$};
\node[parentB] (docb) at (12.265,2.58) {Doc $B$};

\coordinate (rin) at ($(pdfa.east |- render.west)!0.5!(render.west)$);
\draw[black!58,line width=0.52pt] (pdfa.east) -- (rin |- pdfa.east) -- (rin);
\draw[black!58,line width=0.52pt] (pdfb.east) -- (rin |- pdfb.east) -- (rin);
\fill[black!58] (rin) circle (0.55pt);
\draw[flow] (rin) -- (render.west);
\draw[flow] (render.east) -- (2.20,2.83);
\draw[flow] (2.20,2.83) |- (listA.west);
\draw[flow] (2.20,2.83) |- (listB.west);
\draw[ctrlflow] (listA.east) -- (expand.west);
\draw[ctrlflow] (listB.east) -- (expand.west);
\draw[ctrlflow] (expand.east) -- (members.west);
\draw[flow] (members.east) -- (ocr.west);
\draw[flow] (ocr.east) -- (results.west);
\draw[ctrlflow] (results.east) -- (reduce.west);
\draw[ctrlflow] (reduce.east) -- (assemble.west);
\coordinate (aout) at
  ($(assemble.east)!0.5!(doca.west |- assemble.east)$);
\draw[black!58,line width=0.52pt] (assemble.east) -- (aout);
\draw[flow] (aout) |- (doca.west);
\draw[flow] (aout) |- (docb.west);

\draw[lowering] (render.south west) -- (0.225,1.46);
\draw[lowering] (render.south east) -- (3.775,1.46);
\draw[lowering] (ocr.south west) -- (4.075,1.46);
\draw[lowering] (ocr.south east) -- (8.825,1.46);
\draw[lowering] (assemble.south west) -- (9.005,1.46);
\draw[lowering] (assemble.south east) -- (12.555,1.46);

\node[title,text=RayExec!72!black,fill=RayExec!3,inner sep=1.0pt]
  at (0.18,1.73) {Execution plane: one FIFO Ready Queue per Call};
\node[note,anchor=east,text=RayExec!72!black,font=\sffamily\tiny\bfseries,
  fill=RayExec!3,inner sep=0.8pt]
  at (12.66,1.73) {Actors and batches vary; lineage remains fixed.};

\node[operator,minimum width=3.55cm] (rop) at (2.00,0.78) {};
\node[font=\sffamily\scriptsize\bfseries,text=RayExec!76!black,anchor=west]
  at (0.38,1.37) {render Operator};
\node[badge,anchor=east] at (3.62,1.37) {CPU $\times N$};
\node[note,anchor=west,font=\sffamily\tiny\bfseries] at (0.38,1.04) {Ready Queue};
\node[fifo,minimum width=1.08cm] (rfifo) at (1.00,0.70) {};
\node[parentA,minimum width=4.2mm,minimum height=3.8mm] at (0.80,0.70) {$A$};
\node[parentB,minimum width=4.2mm,minimum height=3.8mm] at (1.20,0.70) {$B$};
\node[note] at (1.00,0.43) {FIFO head $\rightarrow$};
\draw[junction] (rfifo.east) -- (1.74,0.70);
\node[actor] (ra1) at (2.55,0.93) {Actor 1 $\cdot$ $\textcolor{RayA}{A}$};
\node[actor] (ra2) at (2.55,0.46) {Actor 2 $\cdot$ $\textcolor{RayB}{B}$};
\draw[branch] (1.74,0.70) |- (ra1.west);
\draw[branch] (1.74,0.70) |- (ra2.west);
\node[note,font=\sffamily\scriptsize] at (3.43,0.70) {$\cdots$};

\node[operator,minimum width=4.75cm] (oop) at (6.45,0.78) {};
\node[font=\sffamily\scriptsize\bfseries,text=RayExec!76!black,anchor=west]
  at (4.25,1.37) {ocr Operator};
\node[badge,anchor=east] at (8.67,1.37) {GPU $\times N$ $\cdot$ batch};
\node[note,anchor=west,font=\sffamily\tiny\bfseries] at (4.28,1.04) {Ready Queue};
\node[fifo,minimum width=1.60cm] (ofifo) at (5.10,0.70) {};
\node[grainA] at (4.50,0.70) {$A_0$};
\node[grainB] at (4.90,0.70) {$B_0$};
\node[grainA] at (5.30,0.70) {$A_1$};
\node[grainB] at (5.70,0.70) {$B_1$};
\node[note] at (5.10,0.43) {FIFO head $\rightarrow$};
\draw[junction] (ofifo.east) -- (6.25,0.70);
\node[actor,minimum width=15.0mm] (oa1) at (7.36,1.10)
  {Actor 1 $\cdot$ $[\textcolor{RayA}{A_0},\textcolor{RayB}{B_0}]$};
\node[actor,minimum width=15.0mm] (oa2) at (7.36,0.70)
  {Actor 2 $\cdot$ $[\textcolor{RayA}{A_1},\textcolor{RayB}{B_1}]$};
\node[actor,minimum width=15.0mm] (oa3) at (7.36,0.30)
  {Actor 3 $\cdot$ $[\textcolor{RayB}{B_2},\textcolor{RayB}{B_3}]$};
\draw[branch] (6.25,0.70) |- (oa1.west);
\draw[branch] (6.25,0.70) -- (oa2.west);
\draw[branch] (6.25,0.70) |- (oa3.west);
\node[note,font=\sffamily\scriptsize] at (8.48,0.70) {$\cdots$};

\node[operator,minimum width=3.55cm] (aop) at (10.78,0.78) {};
\node[font=\sffamily\scriptsize\bfseries,text=RayExec!76!black,anchor=west]
  at (9.18,1.37) {assemble Operator};
\node[badge,anchor=east] at (12.40,1.37) {CPU $\times N$};
\node[note,anchor=west,font=\sffamily\tiny\bfseries] at (9.12,1.04) {Ready Queue};
\node[fifo,minimum width=1.08cm] (afifo) at (9.65,0.70) {};
\node[grainA] at (9.40,0.70) {$R_{A_0}$};
\node[grainB] at (9.90,0.70) {$R_{B_0}$};
\node[note] at (9.65,0.43) {FIFO head $\rightarrow$};
\draw[junction] (afifo.east) -- (10.54,0.70);
\node[actor] (aa1) at (11.45,0.93)
  {Actor 1 $\cdot$ $\textcolor{RayA}{R_{A_0}}$};
\node[actor] (aa2) at (11.45,0.46)
  {Actor 2 $\cdot$ $\textcolor{RayB}{R_{B_0}}$};
\draw[branch] (10.54,0.70) |- (aa1.west);
\draw[branch] (10.54,0.70) |- (aa2.west);
\node[note,font=\sffamily\scriptsize] at (12.33,0.70) {$\cdots$};


\end{tikzpicture}%
}
\end{minipage}
\endgroup
\vspace{-1em}
\caption{From a RayOrch program to lineage-controlled execution.  (a) A
Torch-like program combines ordinary Calls with structural operators.  (b)
Compilation separates structural control (membership and closure) from
execution (per-Call FIFO Ready Queues and actor pools); batching, retries, and placement may
vary without changing the logical structure.}
\Description{A two-part figure.  Panel (a) is a compact, Torch-like Python
pipeline.  Panel (b) shows its corresponding structural-control and execution
planes.  The upper plane shows PDF A and B flowing through render, expand, OCR,
reduction by Expansion, assembly, and ordered output documents.  The lower
plane contains three per-Call Operators.  Each Operator contains a FIFO Ready Queue
followed by multiple actors; actor interiors identify the
parent, page, or reduced-result Grains assigned to that replica.  The OCR
actors carry small cross-parent page batches, while the assemble actors consume
reduced result Items.}
\label{fig:program-control-execution}
\vspace{-1em}
\end{figure*}

%% file: tables/logical_objects.tex
\begin{table*}[t]
\caption{Logical objects used throughout RayOrch, illustrated with the PDF-to-page OCR pipeline.  Their execution realization is described in Section~\ref{sec:design}.}
\vspace{-1em}
\label{tab:logical-objects}
\small
\centering
\setlength{\tabcolsep}{2.5pt}
\renewcommand{\arraystretch}{1.08}
\begin{tabular}{@{}>{\raggedright\arraybackslash}m{0.09\textwidth}|
  >{\raggedright\arraybackslash}m{0.36\textwidth}|
  >{\raggedright\arraybackslash}m{0.12\textwidth}|
  >{\raggedright\arraybackslash}m{0.34\textwidth}@{}}
\hline
\textbf{Object} & \textbf{Direct meaning} & \textbf{Symbol} &
\textbf{Page-OCR example} \\ \hline
\textbf{Domain} & Structural level containing one kind of Entity &
$D$ & \textsc{Page}: all rendered page records \\ \hline
\textbf{Entity} & One data unit with a unique, immutable key $k$ in $D$ &
$e=(D,k)$ & $e_{A_0}=(\textsc{Page},(A,0))$: page 0 of PDF $A$ \\ \hline
\textbf{Expansion} & Ordered child-Entity set materialized for one parent
and child Domain & $X=(C,e_p)$ & Pages materialized from PDF $A$ \\ \hline
\textbf{Function} & Reusable user code wrapped by a \texttt{RayModule} &
$f$ & \texttt{MinerUVlmOcrPage}: reusable page OCR \\ \hline
\textbf{Call} & One configured use of a Function in the program &
$c$ & \texttt{self.ocr}: configured use of that OCR Function \\ \hline
\textbf{Port} & Schedule-independent logical data edge produced by a source,
Call, or structural primitive & $p$ &
$p_{\mathtt{ocr},0}=\texttt{ocr\_results}$: OCR output and
\texttt{F.reduce} input \\ \hline
\textbf{Item} & Value/outcome for one Port-Entity pair; a list remains one
Item & $I=(p,e)$ & $I=(p_{\mathtt{ocr},0},A_0)$: one OCR-result object \\
\hline
\textbf{Grain} & One Call applied to one Entity & $G=(c,e)$ &
\texttt{self.ocr}$(A_0)$: OCR computation for page $A_0$ \\ \hline
\end{tabular}
\end{table*}

%% file: sections/4_design.tex
\section{Design and Execution}
\label{sec:design}

Section~\ref{sec:overview} established the logical dataflow and its separation
from physical execution.  This section follows one declared program through
the runtime: compilation defines its structural relations, lineage state records
what materialized, Ready Queues schedule executable Grains, and commit and
recovery preserve the same logical result despite batching, retries, and worker
replacement.

\subsection{Compiling the Declarative Structural Plan}

Table~\ref{tab:primitives} summarizes the four structural primitives in the
DSL.  The \texttt{F.} prefix distinguishes them from user Functions:
\texttt{F.expand} materializes an ordered child set, and \texttt{F.reduce} is
the matching parent-scoped gather; \texttt{F.filter} changes membership within
a Domain, while \texttt{F.broadcast} makes an ancestor Item available in a
descendant Domain.  Aligned variants preserve member alignment, and optional
inputs affect propagation without adding a structural node.  Together these
operations express hierarchical fan-out without arbitrary joins or regrouping. 
The compiler lowers the symbolic DSL into an immutable graph after checking
acyclicity, Domain compatibility, and each declared pair.  The graph
contains Calls, Ports, Domains, consumers, and pre-indexed structural effects;
invalid cross-Domain uses fail before execution.  The runtime therefore
consumes declared relations instead of inferring lineage from payloads or
user-managed identifiers.  The graph fixes membership and consumer rules, but
the concrete child Entities and Grains remain runtime-produced.

\input{tables/primitives}

\subsection{Lineage State and Parent Completion}

For each child Domain $C$ and parent Entity $e_p$, the runtime creates one
\emph{Expansion} $X=(C,e_p)$ containing the ordered child Entities materialized
for that parent.  Each logical object publishes exactly one terminal fact:
\begin{equation}
\begin{aligned}
\mathsf{Entity} &: \mathsf{Published}(e,o),\\
\mathsf{Item} &: \mathsf{Present}(v)\mid\mathsf{Dropped}
       \mid\mathsf{Failed}\mid\mathsf{Suppressed},\\
\mathsf{Expansion} &: \mathsf{Succeeded}(\langle e_0,\ldots,e_{m-1}\rangle)
       \mid\mathsf{Dropped}\mid\mathsf{Failed}.
\end{aligned}
\label{eq:flat-facts}
\end{equation}
Here $o$ is the parent's immutable ordinal.  Entity facts record existence and
lineage; \textsc{failed} is a direct producer failure, whereas \textsc{suppressed}
means an input prevents downstream work.  An empty successful Expansion
$\langle\rangle$ is terminal.  These logical facts are separate from a Grain's
physical phase.  Publications are irreversible: identical repeats are
idempotent and conflicts are rejected.  The lineage engine alone publishes
them; the dispatcher owns Grain phases and attempt generations, and executors
own capacity and RPCs.

Input propagation follows a fixed precedence: failed or suppressed input
suppresses outputs; unresolved input waits; a dropped required input drops the
output; otherwise the Grain becomes ready, with a dropped optional input
exposed as missing.  Filter and broadcast follow the same propagation rule.

The Reduce operation emits one parent Item only after its Expansion, every
child-membership outcome, and every surviving value resolve.  Dropped members
are excluded; failed or suppressed members suppress the parent; present values
are emitted in immutable ordinal order.  Zero survivors yield a present empty
list.  A physical batch can therefore never define a partial document: it only
determines which ready Grains share an attempt, not the reduction relation.

\subsection{Scheduling Ready Grains}
\label{sec:design-scheduling}

For each input microbatch, each Call owns one per-Call FIFO Ready Queue for
READY Grains.  It is an admission queue, not a grouping operation: a Grain
enters once its inputs are ready, and downstream Calls receive entries only
after an upstream outcome commits.
Figure~\ref{fig:runtime-fifo-pipeline} shows the normal path: one
committed outcome can activate multiple downstream Calls, whose Ready Queues
dispatch independently rather than waiting for a stage-wide regrouping barrier.
Recovery uses separate priority queues that reinsert exact Grain groups without
changing normal Ready-Queue order.

\input{figures/runtime_fifo_pipeline}

Resources, replicas, and batching scope are configured per Call.
Elastic reservation removes at most $B$ entries from the Ready Queue in $O(B)$ and may mix
parents.  Parent-bound reservation takes up to $B$ siblings of the first ready
parent while preserving the relative order of all remaining queue entries.  Both policies alter only the
physical batch; Grain identity, membership, and reconstruction order remain
unchanged.  A completed parent can thus enable its next-stage work while
another parent is still producing siblings.

Both policies share one physical Grain lifecycle.  Here \textsc{waiting} means
that inputs are not yet resolved, \textsc{ready} means executable, and
\textsc{sealed} means that a logical terminal outcome has been accepted:
\begin{equation}
\begin{aligned}
\textsc{waiting}&\xrightarrow{\text{ready}}\textsc{ready}
\xrightarrow{\text{reserve}}\textsc{in-flight}
\xrightarrow{\text{commit}}\textsc{sealed},\\[-0.2em]
\textsc{in-flight}&\xrightarrow{\text{retry}}\textsc{ready},\qquad
\textsc{waiting}\xrightarrow{\text{terminal}}\textsc{sealed}.
\end{aligned}
\label{eq:grain-phase}
\end{equation}

The reservation policy chooses which ready Grains share an attempt; it does
not change this lifecycle or the logical outcome.

\subsection{Commit, Failure, and Recovery}
\label{sec:design-recovery}

\input{figures/failure_scope_minimal}

Figure~\ref{fig:recovery-protocol} separates logical commit from physical
attempts.  A typed \texttt{RecordFailure} is terminal for one Grain.  A typed
\texttt{GroupFailure} installs a local suppression barrier for one
$(\text{Call},\text{parent})$: queued siblings are suppressed before dispatch,
in-flight siblings may finish but cannot commit, and already committed
siblings remain untouched.  Unrelated parents and queues remain live.

An untyped UDF exception or infrastructure failure is instead a physical
attempt failure.  The runtime may retry the Grain or replace its actor and
replay the Grain at a higher generation; these actions preserve Grain identity
and change only the physical attempt.

A report $r$ commits only if its Grain is still in flight and its generation
(the attempt number) is current:
\begin{equation}
\begin{aligned}
\mathsf{Accept}(r,G)\ \Longleftrightarrow\quad
 &\mathsf{phase}(G)=\mathsf{IN\mbox{-}FLIGHT}\\[-0.15em]
 &{}\land\ r.\mathsf{grain}=G\\[-0.15em]
 &{}\land\ r.\mathsf{generation}=\mathsf{generation}(G).
\end{aligned}
\label{eq:generation-fence}
\end{equation}
An accepted report seals the Grain and publishes terminal facts atomically; the
generation test rejects stale or duplicate reports.

\subsection{Semantic Contract}
\label{sec:design-invariance}

For a verified finite, acyclic hierarchical $1{:}M$ program $P$ and fixed
per-Grain outcomes $\Omega$, let $S_{P,\Omega}(\sigma)$ denote the final Items
and their lineage order under legal physical schedule $\sigma$.  A schedule
may change batch composition, worker placement, or retry timing; the runtime
contract is:
\begin{equation}
 S_{P,\Omega}(\sigma_1)=S_{P,\Omega}(\sigma_2).
\label{eq:schedule-invariance}
\end{equation}

The equality follows from four invariants: batch-independent Grain identity,
unique monotone lineage facts, generation-fenced retries, and ordinal-ordered
Reduce.  Thus ownership, completion, and reconstruction order do not depend on
the physical schedule.  If a UDF is also invariant to batch shape, input order,
and randomness, the guarantee extends to payload bytes; otherwise it covers
structure and terminal status only.

%% file: tables/primitives.tex
\begin{table}[H]
\caption{RayOrch's structural primitives in program notation.  Arguments and
results are Ports.}
\vspace{-1em}
\label{tab:primitives}
\small
\centering
\setlength{\tabcolsep}{2.5pt}
\renewcommand{\arraystretch}{1.08}
\begin{tabular}{@{}>{\raggedright\arraybackslash}m{0.33\columnwidth}|
  >{\centering\arraybackslash}m{0.22\columnwidth}|
  >{\centering\arraybackslash}m{0.39\columnwidth}@{}}
\hline
\textbf{Primitive} & \textbf{Domain effect} & \textbf{Direct meaning} \\
\hline
\textbf{\texttt{F.expand(x)}} & parent $\to$ child &
List elements become ordered child Items \\ \hline
\textbf{\texttt{F.filter(x, mask)}} & same Domain &
Pass $x$ on true; drop the output on false \\ \hline
\textbf{\texttt{F.broadcast(x, like=y)}} &
\shortstack{ancestor\\$\to$ descendant} &
Make ancestor $x$ available in $y$'s Domain \\ \hline
\textbf{\texttt{F.reduce(x)}} & child $\to$ parent &
Surviving Items form one ordered parent list \\ \hline
\end{tabular}
\end{table}

%% file: figures/runtime_fifo_pipeline.tex
\begin{figure*}[t]
\centering
\resizebox{1.0\textwidth}{!}{%
\begin{tikzpicture}[
  font=\sffamily\footnotesize,
  >={Latex[length=1.45mm,width=0.95mm]},
  panel/.style={draw=black!24, fill=black!1, rounded corners=3pt,
    line width=0.55pt},
  driverpanel/.style={draw=OneDarkBlue!68!black, dashed,
    line width=0.65pt, rounded corners=3pt, fill=OneDarkBlue!3},
  dataplane/.style={draw=OneDarkGreen!62!black, dashed,
    line width=0.65pt, rounded corners=3pt, fill=OneDarkGreen!3},
  box/.style={draw=black!52, line width=0.5pt, rounded corners=2pt,
    fill=white, minimum height=7.5mm, align=center, inner sep=2.2pt},
  queue/.style={box, fill=black!1},
  actor/.style={box, draw=OneDarkGreen!62!black, fill=OneDarkGreen!10},
  semantic/.style={box, fill=OneDarkBlue!9},
  fact/.style={box, draw=OneDarkPurple!68!black, fill=OneDarkPurple!9},
  callcard/.style={draw=OneDarkPurple!58!black, line width=0.55pt,
    rounded corners=2pt, fill=white},
  flow/.style={->, semithick, draw=black!78},
  semanticflow/.style={->, semithick, draw=OneDarkBlue!72!black},
  zoomlink/.style={draw=OneDarkPurple!72!black, densely dashed,
    line width=0.65pt},
  title/.style={font=\sffamily\small\bfseries, anchor=west},
  planelabel/.style={font=\sffamily\scriptsize\bfseries, anchor=west},
  edgelabel/.style={font=\sffamily\scriptsize, fill=white, inner sep=0.6pt},
  transitionlabel/.style={font=\sffamily\scriptsize, inner sep=0pt},
  miniq/.style={queue, minimum width=1.15cm, minimum height=5.8mm,
    inner sep=1.0pt, font=\sffamily\scriptsize\bfseries},
  minipool/.style={actor, minimum width=1.15cm, minimum height=5.8mm,
    inner sep=1.0pt, font=\sffamily\scriptsize\bfseries},
  callflow/.style={callcard, minimum width=2.50cm, minimum height=8.4mm,
    align=center, inner sep=1.6pt, font=\sffamily\scriptsize},
  branchqueue/.style={queue, minimum width=1.85cm, minimum height=7.5mm,
    inner sep=1.4pt, font=\sffamily\scriptsize},
  grain/.style={draw, rounded corners=1pt, minimum width=4.6mm,
    minimum height=5.0mm, inner sep=1pt, font=\sffamily\scriptsize\bfseries},
  minigrain/.style={draw, rounded corners=1pt, minimum width=4.2mm,
    minimum height=4.4mm, inner sep=0.7pt,
    font=\sffamily\scriptsize\bfseries}
]

\draw[panel] (0.00,0.00) rectangle (5.05,4.55);
\draw[panel] (5.40,0.00) rectangle (17.75,4.55);
\node[title] at (0.18,4.30) {(a) A fork in the Call DAG};
\node[title] at (5.62,4.30) {(b) One outcome can ready multiple Calls};

\draw[zoomlink] (0.28,0.30) rectangle (4.77,4.00);
\node[font=\sffamily\scriptsize\bfseries,
  text=OneDarkPurple!78!black, fill=white, inner sep=1.0pt]
  at (2.52,4.00) {selected subgraph};

\node[queue, minimum width=1.25cm, minimum height=5.8mm]
  (groupA) at (0.98,3.48) {};
\node[minigrain, draw=OneDarkBlue!90!black, text=OneDarkBlue!75!black,
  fill=OneDarkBlue!5] at (0.69,3.48) {$A_0$};
\node[minigrain, draw=OneDarkBlue!90!black, text=OneDarkBlue!75!black,
  fill=OneDarkBlue!5] at (1.27,3.48) {$A_1$};

\node[queue, minimum width=1.25cm, minimum height=5.8mm]
  (groupB) at (2.52,3.48) {};
\node[minigrain, draw=OneDarkRed!90!black, text=OneDarkRed!80!black,
  fill=OneDarkRed!5] at (2.23,3.48) {$B_0$};
\node[minigrain, draw=OneDarkRed!90!black, text=OneDarkRed!80!black,
  fill=OneDarkRed!5] at (2.81,3.48) {$B_1$};

\node[queue, minimum width=1.25cm, minimum height=5.8mm]
  (groupC) at (4.06,3.48) {};
\node[minigrain, draw=OneDarkGreen!65!black, text=OneDarkGreen!45!black,
  fill=OneDarkGreen!5] at (3.77,3.48) {$C_0$};
\node[minigrain, draw=OneDarkGreen!65!black, text=OneDarkGreen!45!black,
  fill=OneDarkGreen!5] at (4.35,3.48) {$C_1$};

\node[callflow, minimum width=2.68cm] (callCOverview) at (2.52,2.18)
  {\textbf{Call $c$}\\[-1pt]FIFO Ready Queue\\[-1pt]$\longrightarrow$ actor pool};
  \coordinate (sourceBusLeft) at (0.98,2.94);
  \coordinate (sourceBusMid) at (2.52,2.94);
  \coordinate (sourceBusRight) at (4.06,2.94);
\draw[semithick, draw=black!78] (groupA.south) -- (sourceBusLeft);
\draw[semithick, draw=black!78] (groupB.south) -- (sourceBusMid);
\draw[semithick, draw=black!78] (groupC.south) -- (sourceBusRight);
\draw[semithick, draw=black!78] (sourceBusLeft) -- (sourceBusRight);
\draw[flow] (sourceBusMid) -- (callCOverview.north);

\node[callflow, minimum width=1.84cm] (callDOverview) at (1.35,0.83)
  {\textbf{Call $d$}\\[-1pt]FIFO Ready Queue\\[-1pt]$\longrightarrow$ pool};
\node[callflow, minimum width=1.84cm] (callEOverview) at (3.69,0.83)
  {\textbf{Call $e$}\\[-1pt]FIFO Ready Queue\\[-1pt]$\longrightarrow$ pool};

\coordinate (forkOverview) at (2.52,1.52);
\draw[semithick, draw=black!78] (callCOverview.south) -- (forkOverview);
\draw[flow] (forkOverview) -| (callDOverview.north);
\draw[flow] (forkOverview) -| (callEOverview.north);
\fill[black!72] (forkOverview) circle (0.7pt);

\draw[driverpanel] (5.65,1.70) rectangle (17.50,4.02);
\node[planelabel, text=OneDarkBlue!72!black] at (5.90,3.80)
  {RayOrch driver (control plane)};
\node[semantic, minimum width=10.95cm, minimum height=6.1mm]
  (lineage) at (11.58,3.38)
  {\textbf{persistent structural lineage}: parent $\cdot$ ordinal $\cdot$
   membership $\cdot$ terminal outcome};

\node[queue, minimum width=2.15cm] (qc) at (6.75,2.38) {};
\node[font=\sffamily\scriptsize\bfseries, anchor=west]
  at (5.73,2.59) {\texttt{Ready Queue[c]}};
\node[grain, draw=OneDarkBlue!90!black, text=OneDarkBlue!75!black]
  at (6.13,2.25) {$A_1$};
\node[grain, draw=OneDarkRed!90!black, text=OneDarkRed!80!black]
  at (6.70,2.25) {$B_1$};
\node[grain, draw=OneDarkGreen!65!black, text=OneDarkGreen!45!black]
  at (7.27,2.25) {$C_1$};
\node[font=\sffamily\scriptsize, text=black!58] at (7.68,2.25) {$\cdots$};

\node[box, fill=OneDarkBlue!8, minimum width=1.05cm, minimum height=7.5mm]
  (gate) at (8.50,2.38) {per-Grain\\commit};
\node[fact, minimum width=1.20cm, minimum height=7.5mm]
  (facts) at (9.85,2.38) {committed\\outcome};
\node[box, minimum width=1.85cm, minimum height=7.5mm]
  (rules) at (11.55,2.38) {input\\propagation rules};

\node[branchqueue] (qd) at (13.95,2.38) {};
\node[font=\sffamily\scriptsize\bfseries, align=center]
  at (13.51,2.38) {\texttt{Ready}\\[-1pt]\texttt{Queue[d]}};
\node[minigrain, draw=OneDarkBlue!90!black, text=OneDarkBlue!75!black,
  fill=OneDarkBlue!5] at (14.16,2.38) {$A_1$};
\node[minigrain, draw=OneDarkRed!90!black, text=OneDarkRed!80!black,
  fill=OneDarkRed!5] at (14.62,2.38) {$B_1$};

\node[branchqueue] (qe) at (16.30,2.38) {};
\node[font=\sffamily\scriptsize\bfseries, align=center]
  at (15.86,2.38) {\texttt{Ready}\\[-1pt]\texttt{Queue[e]}};
\node[minigrain, draw=OneDarkBlue!90!black, text=OneDarkBlue!75!black,
  fill=OneDarkBlue!5] at (16.51,2.38) {$A_1$};
\node[minigrain, draw=OneDarkGreen!65!black, text=OneDarkGreen!45!black,
  fill=OneDarkGreen!5] at (16.97,2.38) {$C_1$};

\draw[flow] (gate.east) -- (facts.west);
\node[transitionlabel] at (9.18,2.83) {accept};
\draw[semanticflow] (facts.east) -- (rules.west);
\node[transitionlabel] at (10.70,2.83) {advance};
\coordinate (fork) at (12.75,2.38);
\coordinate (busTurn) at (12.75,2.94);
\coordinate (busD) at (13.95,2.94);
\coordinate (busE) at (16.30,2.94);
\draw[semithick, draw=black!78] (rules.east) -- (fork) --
  (busTurn) -- (busE);
\draw[flow] (busD) -- (qd.north);
\draw[flow] (busE) -- (qe.north);
\draw[semanticflow] (gate.north) -- (gate.north |- lineage.south);

\draw[dataplane] (5.65,0.14) rectangle (17.50,1.50);
\node[font=\sffamily\scriptsize\bfseries, fill=white, inner xsep=2.5pt,
  text=OneDarkGreen!52!black] at (10.55,1.50)
  {Ray actor data plane};

\node[actor, minimum width=3.00cm] (actorC) at (7.20,0.78)
  {\textbf{Call $c$ actor pool}\\[-1pt]
   {\scriptsize\texttt{batch\_size[c]}}\\[-1pt]
   $[A_1,B_1,C_1] \rightarrow [A_2,B_2]$};
\node[actor, minimum width=3.00cm] (actorD) at (11.58,0.78)
  {\textbf{Call $d$ actor pool}\\[-1pt]
   {\scriptsize\texttt{batch\_size[d]}}\\[-1pt]
   starts independently};
\node[actor, minimum width=3.00cm] (actorE) at (15.96,0.78)
  {\textbf{Call $e$ actor pool}\\[-1pt]
   {\scriptsize\texttt{batch\_size[e]}}\\[-1pt]
   starts independently};

\draw[flow] (qc.south) -- node[edgelabel, left, pos=0.33] {dispatch}
  (qc.south |- actorC.north);
\draw[flow] (actorC.north east) -- node[edgelabel, right] {return}
  (gate.south);
\draw[flow] (qd.south) -- node[edgelabel, left, pos=0.33] {dispatch}
  (actorD.north);
\draw[flow] (qe.south) -- node[edgelabel, right, pos=0.33] {dispatch}
  (actorE.north);

\end{tikzpicture}%
}
\vspace{-1em}
\caption{A committed outcome can activate multiple downstream Calls.  (a) Call
$c$ feeds consumers $d$ and $e$, each with its own FIFO Ready Queue and actor pool.
(b) The driver propagates the committed Grain to both Ready Queues, which dispatch
independently under their configured batch sizes without a stage-wide
regrouping barrier.}
\Description{A two-panel runtime diagram.  The left panel shows upstream Grain
groups A, B, and C flowing into Call c, which forks to Calls d and e.  Each Call contains
a FIFO Ready Queue paired with an actor pool.  The right panel enlarges the fork.  In
the driver control plane, Call c's FIFO Ready Queue dispatches to its actor pool, a
per-Grain commit accepts the report, and the committed outcome reaches input
propagation rules.  The rules branch to separate FIFO Ready Queues for Calls d and e, so
each consumer can advance independently.  In the Ray actor data plane, those
queues dispatch work to independent actor pools d and e while Call c can
continue with another batch.}
\label{fig:runtime-fifo-pipeline}
\vspace{-1em}
\end{figure*}
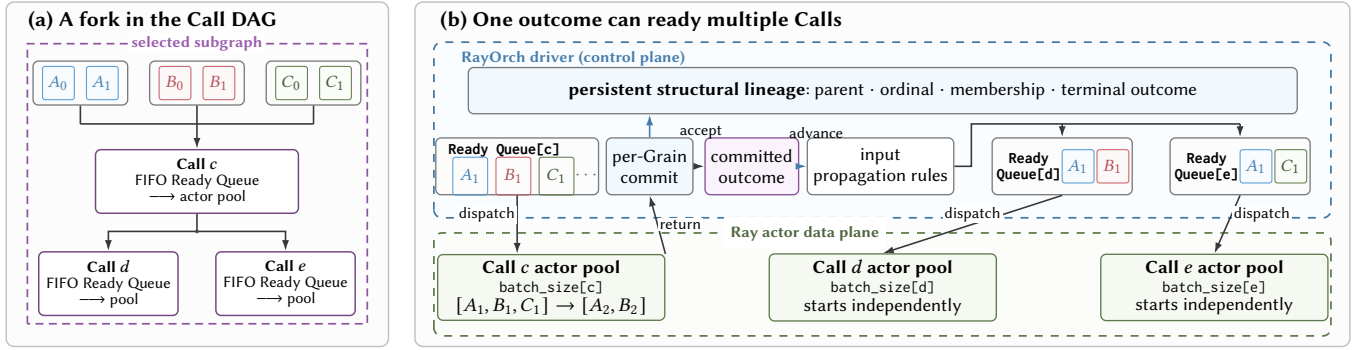

%% file: figures/failure_scope_minimal.tex
\begin{figure}[t]
\centering
\begingroup
\resizebox{0.95\columnwidth}{!}{%
\begin{tikzpicture}[
  font=\sffamily\scriptsize,
  >={Latex[length=1.5mm,width=0.95mm]},
  flow/.style={->, semithick, draw=black!62},
  semantic/.style={->, semithick, draw=OneDarkPurple!72!black},
  queue/.style={draw=black!43, rounded corners=1.6pt, fill=black!1,
    minimum width=5.90cm, minimum height=6.2mm},
  actor/.style={draw=black!48, dashed, rounded corners=1.6pt, fill=black!1,
    minimum width=3.55cm, minimum height=6.2mm},
  outcome/.style={dashed, rounded corners=1.6pt, minimum height=9.0mm},
  grain/.style={draw, rounded corners=1.0pt, minimum width=4.7mm,
    minimum height=4.7mm, inner sep=0.4pt, font=\sffamily\tiny},
  agrain/.style={grain, draw=OneDarkBlue!85!black,
    text=OneDarkBlue!72!black, fill=OneDarkBlue!9},
  bgrain/.style={grain, draw=OneDarkRed!80!black,
    text=OneDarkRed!82!black, fill=OneDarkRed!9},
  cgrain/.style={grain, draw=OneDarkGreen!62!black,
    text=OneDarkGreen!48!black, fill=OneDarkGreen!9},
  rowlabel/.style={font=\sffamily\scriptsize\bfseries, anchor=east},
  label/.style={font=\sffamily\tiny, text=black!62}
]

\node[rowlabel] at (1.22,3.08) {\texttt{Ready Queue[c]}};
\node[queue] (queue) at (4.35,3.08) {};
\node[agrain] at (1.82,3.08) {$A_1$};
\node[bgrain] at (2.53,3.08) {$B_1$};
\node[cgrain] at (3.24,3.08) {$C_1$};
\node[bgrain] at (3.95,3.08) {$B_2$};
\draw[black!36] (4.35,2.82) -- (4.35,3.34);
\node[bgrain] (qb3) at (4.85,3.08) {$B_3$};
\node[label, anchor=west, text=OneDarkRed!72!black]
  at (5.23,3.08) {still \textsc{ready}};

\node[rowlabel] at (1.22,2.12) {ACTOR};
\node[actor] (actor) at (3.12,2.12) {};
\node[agrain] at (1.82,2.12) {$A_1$};
\node[bgrain] at (2.53,2.12) {$B_1$};
\node[cgrain] at (3.24,2.12) {$C_1$};
\node[bgrain] at (3.95,2.12) {$B_2$};
\draw[flow] (3.12,2.76) -- node[right, label] {reserve 4} (3.12,2.45);
\node[label, anchor=west] at (4.98,2.12) {mixed-parent batch};

\node[rowlabel] at (1.22,0.72) {COMMIT};
\draw[OneDarkPurple!72!black, line width=1.15pt] (1.32,1.45) -- (7.30,1.45);
\node[label, anchor=west, text=OneDarkPurple!72!black] at (3.72,1.60)
  {commit gate};
\draw[flow] (actor.south) -- node[right, label] {reports} (3.12,1.48);

\node[outcome, draw=OneDarkGreen!58!black, minimum width=2.05cm,
  minimum height=10.5mm]
  (okgroup) at (2.55,0.72) {};
\node[agrain] at (2.13,0.82) {$A_1$};
\node[cgrain] at (2.96,0.82) {$C_1$};
\node[font=\sffamily\scriptsize, text=OneDarkGreen!52!black]
  at (2.13,0.49) {\ding{51}};
\node[font=\sffamily\scriptsize, text=OneDarkGreen!52!black]
  at (2.96,0.49) {\ding{51}};
\node[label, text=OneDarkGreen!48!black] at (2.55,0.31)
  {commit independently};

\node[outcome, draw=OneDarkRed!76!black, minimum width=3.20cm,
  minimum height=10.5mm]
  (bgroup) at (5.70,0.72) {};
\node[bgrain] at (4.62,0.82) {$B_1$};
\node[bgrain] at (5.70,0.82) {$B_2$};
\node[bgrain] (ob3) at (6.78,0.82) {$B_3$};
\node[font=\sffamily\scriptsize, text=OneDarkRed!82!black]
  at (4.62,0.49) {\ding{55}};
\node[font=\sffamily\scriptsize, text=OneDarkRed!82!black]
  at (5.70,0.49) {\ding{55}};
\node[font=\sffamily\scriptsize, text=OneDarkRed!82!black]
  at (6.78,0.49) {\ding{55}};
\node[label, text=OneDarkRed!82!black] at (4.62,0.31) {group fail};
\node[label, text=OneDarkRed!82!black] at (5.70,0.31) {reject};
\node[label, text=OneDarkRed!82!black] at (6.78,0.31) {suppress};

\draw[OneDarkGreen!52!black, line width=0.65pt]
  (1.72,0.04) -- (3.38,0.04);
\draw[OneDarkGreen!52!black, line width=0.65pt]
  (1.72,0.04) -- (1.72,0.13)
  (3.38,0.04) -- (3.38,0.13);
\node[label, font=\sffamily\tiny\bfseries, text=OneDarkGreen!48!black]
  at (2.55,-0.15) {PASS: parents $A,C$};

\draw[OneDarkRed!76!black, line width=0.65pt]
  (4.44,0.04) -- (6.96,0.04);
\draw[OneDarkRed!76!black, line width=0.65pt]
  (4.44,0.04) -- (4.44,0.13)
  (6.96,0.04) -- (6.96,0.13);
\node[label, font=\sffamily\tiny\bfseries, text=OneDarkRed!82!black]
  at (5.70,-0.15) {REJECT: parent $B$};

\draw[semantic] (qb3.south) -- (4.85,2.66) --
  (7.42,2.66) -- (7.42,0.82) -- (ob3.east);
\node[label, rotate=90, fill=white, inner sep=0.7pt,
  text=OneDarkPurple!72!black] at (7.42,1.84) {never enters actor};

\draw[flow] (2.55,1.42) -- (okgroup.north);
\draw[semantic] (5.70,1.42) -- (bgroup.north);

\end{tikzpicture}%
}
\endgroup
\vspace{-1em}
\caption{Actor execution and commit have different scopes.  A failure in parent
$B$ rejects its in-flight report and suppresses queued siblings, while committed
work and unrelated parents remain unaffected.}
\Description{A compact three-row diagram.  The FIFO Ready Queue for Call c contains A1, B1,
C1, B2, and B3.  The actor row contains only reserved A1, B1, C1, and B2.  A
thin semantic commit gate separates actor reports from terminal effects.  A1
and C1 commit independently, so parents A and C pass.  B1 reports an explicit
GroupFailure for parent B; B2's in-flight report is rejected at commit, and B3
is suppressed in the FIFO without entering the actor.  Already committed work
is unaffected, and the rejected sibling outcomes form a lineage-scoped parent-B
barrier.}
\label{fig:recovery-protocol}
\vspace{-1em}
\end{figure}
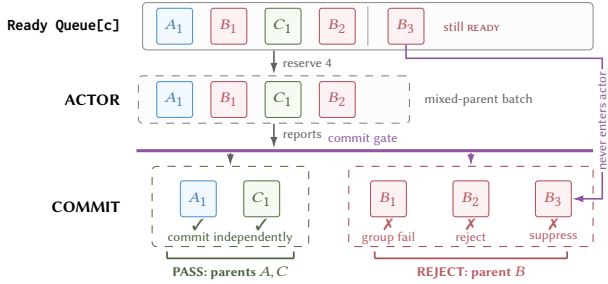

%% file: sections/5_implementation.tex
\section{Implementation}
\label{sec:implementation}

\sys is implemented as a Python layer on Ray~\cite{moritz2018ray}.  Figure~\ref{fig:runtime-fifo-pipeline}
shows the resulting mapping: each configured Call is backed by a persistent Ray
actor pool, while the driver retains the compiled plan, lineage metadata, and
pending \texttt{ObjectRef}s.  Payloads remain in Ray's object store; actor
handles and references are physical execution state and never enter a Grain's
logical identity.

Figure~\ref{fig:program-control-execution} also makes the stage interface
explicit.  \texttt{F.*} denotes \sys's built-in structural operators, with
\texttt{F.expand} and \texttt{F.reduce} expressing logical grain/domain
changes.  Each model stage is a Python class wrapped by \texttt{RayModule} and
implements a batch UDF of the form
$\mathrm{List}[\mathrm{Object}]\!\to\!\mathrm{List}[\mathrm{Object}]$;
for $n$ input Grains, the lists contain $n$ corresponding values, while an
expanded row may itself be a list.  The worker validates the declared arity and
lengths, stores output columns, and returns one per-Grain report per output
Port.  Here \texttt{Object} denotes a Port payload, not a Ray
\texttt{ObjectRef}; UDFs therefore do not maintain parent IDs or ordering.

The stage lifecycle follows the Torch module pattern: \texttt{\_\_init\_\_}
loads models and other heavyweight resources, whereas \texttt{run} performs
the batch computation.  Persistent actor instances amortize initialization
across batches.  Our adapters cover three workload families: MinerU's
rendering, page OCR, and assembly stages (page OCR uses MinerU's 1.2B-parameter
VLM); Docling's layout, OCR, and table stages; and video pipelines that apply
Qwen2.5-VL-7B after decoding clips, frames, or audio and before per-source
merge.  They exchange workload values only; \sys provides lineage, batching,
completion, and the instrumentation used in Section~\ref{sec:evaluation}.

%% file: sections/6_evaluation.tex
\section{Evaluation}
\label{sec:evaluation}

We design the evaluation to answer four key research questions:
\begin{itemize}[leftmargin=*]
  \setlength{\itemsep}{0pt}
  \setlength{\parsep}{0pt}
  \setlength{\topsep}{2pt}
  \item \textbf{RQ1:} How do \sys's end-to-end performance and scalability
  compare with the baselines?
  \item \textbf{RQ2:} What explains \sys's end-to-end gains?
  \item \textbf{RQ3:} How can FIFO scheduling perform beyond
  1:$M$ rebatching?
  \item \textbf{RQ4:} Can runtime-owned lineage suppress doomed sibling work
  while preserving all expected healthy outputs?
\end{itemize}

\vspace{-1.0em}
\subsection{Experimental Setup}
\label{sec:eval-setup}

\noindent\textbf{Hardware and software.}
All experiments run on NVIDIA H20 GPUs: MinerU uses 4--64, video uses 8--64,
and Docling, the ablation, and failure containment use 4 GPUs.
Table~\ref{tab:eval-setup} reports the software and model versions.

\vspace{0.5em}
\noindent\textbf{Workloads and baselines.}
Table~\ref{tab:eval-setup} lists five workloads.  MinerU has 3,689 PDFs and
174,744 valid pages (1--427 pages per PDF; one unreadable PDF is excluded),
video has 27,091 videos and 104,952 clips, and Docling has 2,000 PDFs.  MinerU
uses Ray Data, Daft, and native MinerU as baselines; video uses Ray Data and
Daft; and Docling uses Ray Data and Docling Serve.  Systems use the same valid
inputs and model pipeline within each workload.

\input{tables/evaluation_setup}

\vspace{0.5em}
\noindent\textbf{Metrics and protocol.}
We report E2E wall time, throughput, strong-scaling speedup, and elapsed stage
windows.  Speedups use unrounded measurements; displayed hours are rounded.
Stage windows overlap and are not additive.  The scaling sweeps and cold-run
breakdown come from separate run series.  The ablation reports wall time
relative to the full system; the failure study reports suppressed siblings,
non-trigger UDF calls, and paired time reduction over three runtimes.  We
separately check identity, order, completeness, and failure outcomes.
\emph{Grain} denotes the runtime execution unit.

\vspace{-1.0em}
\subsection{RQ1: Performance and Scalability}
\label{sec:eval-scalability}

\begin{figure}[t]
  \centering
  \includegraphics[width=\columnwidth]{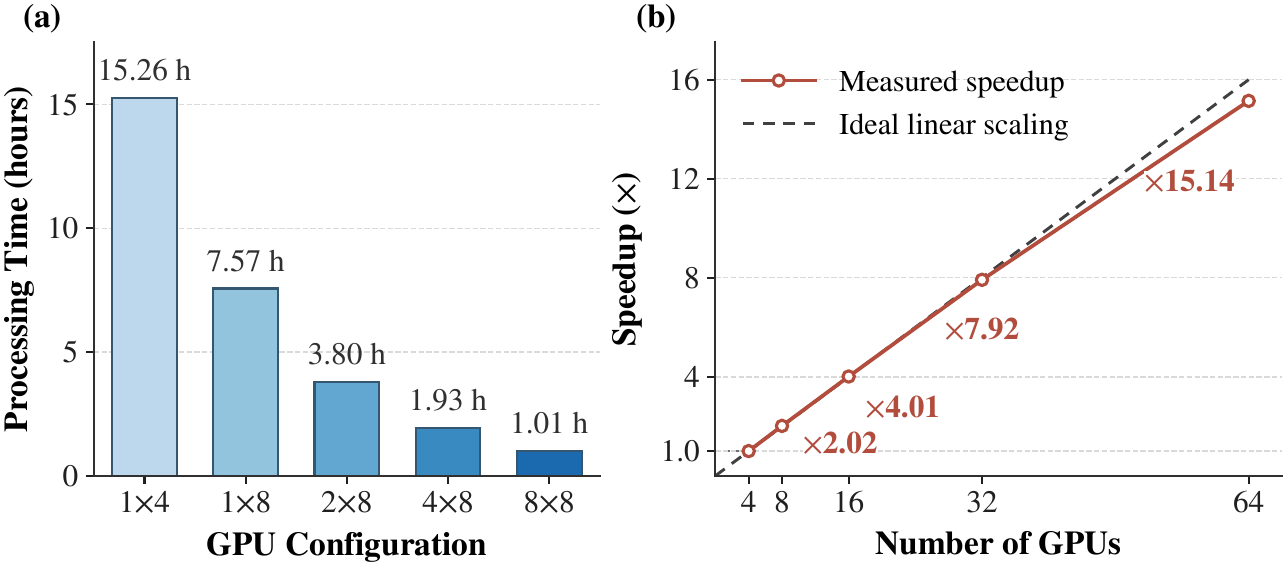}
  \vspace{-1em}
  \caption{Strong scalability of \sys on MinerU.}
  \Description{Two plots show strong scaling from 4 to 64 GPUs.  Processing
  time falls from 15.26 hours to 1.01 hours, while measured speedup rises from
  1.0 to 15.14 and remains close to the ideal linear-scaling line.}
  \label{fig:scale-up}
\end{figure}

\begin{figure}[t]
  \centering
  \includegraphics[width=\columnwidth]{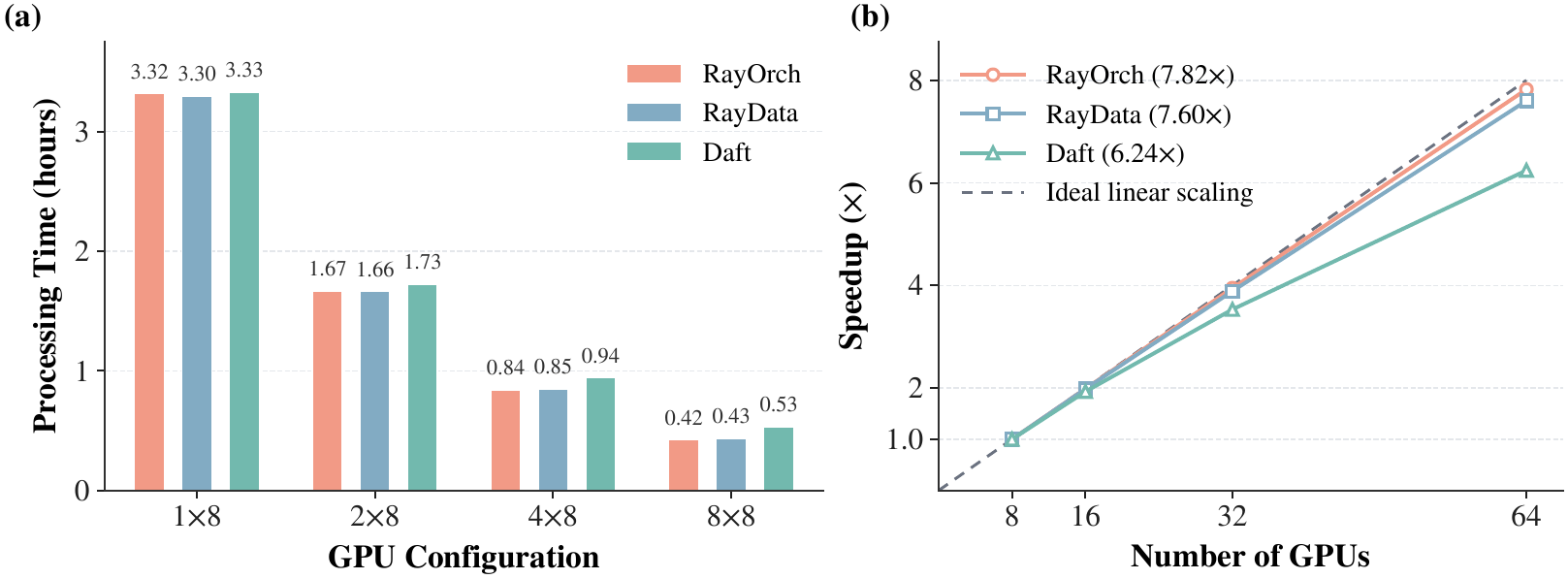}
  \vspace{-1.0em}
  \caption{Scale-up comparison on the complete video workload from
  8 to 64 GPUs.}
  \Description{Two plots compare \sys, Ray Data, and Daft as the allocation
  grows from 8 to 64 GPUs.  \sys processing time falls from 3.32 hours to
  0.42 hours and reaches 7.82 times speedup; Ray Data reaches 7.60 times
  speedup, and Daft reaches 6.24 times speedup.}
  \label{fig:video-scale}
  \vspace{-1.0em}
\end{figure}

\vspace{0.5em}
\noindent\textbf{\sys combines high throughput with near-linear scaling.}
RQ1 pairs the fixed-input strong-scaling sweeps in
Figures~\ref{fig:scale-up} and~\ref{fig:video-scale} with the fixed-allocation
comparisons in Tables~\ref{tab:mineru-e2e} and~\ref{tab:docling-e2e}.  On
MinerU, processing time falls from 15.26 hours on 4 GPUs to 1.01 hours on 64
GPUs; the intermediate $2.02\times$, $4.01\times$, and $7.92\times$ speedups
at 8, 16, and 32 GPUs lead to $15.14\times$ at 64 GPUs, or 94.6\% of ideal
linear scaling.  In the separate 64-GPU E2E comparison, \sys processes
174,744 valid pages in 4,295.7 seconds at 40.6788 pages/s, reducing wall time
by 13.1\%, 29.0\%, and 51.6\% versus Ray Data, Daft, and native MinerU; its
throughput is correspondingly 15.1\%, 40.8\%, and 106.6\% higher.  The video
systems begin at near parity on 8 GPUs, taking 3.32, 3.30, and 3.33 hours for
\sys, Ray Data, and Daft, but diverge as the allocation grows: at 64 GPUs,
\sys finishes in 0.42 hours with $7.82\times$ speedup, compared with 0.43
hours and $7.60\times$ for Ray Data and 0.53 hours and $6.24\times$ for Daft.
On the 2,000-PDF Docling workload, \sys finishes on 4 GPUs in 9,489.07 seconds at
0.2107 documents/s, reducing E2E time by 16.0\% versus Ray Data and 22.3\%
versus Docling Serve.

\input{tables/mineru_e2e}

\input{tables/docling_e2e}

\vspace{0.5em}
\noindent
{\setlength{\fboxsep}{4pt}%
\setlength{\fboxrule}{0.5pt}%
\fcolorbox{blue!45}{blue!5}{%
  \parbox{\dimexpr\columnwidth-2\fboxsep-2\fboxrule\relax}{%
    \textbf{Answer to RQ1.} \sys sustains near-linear scaling through 64
    GPUs, leads both tested document stacks at fixed allocations, and delivers
    the fastest 64-GPU video result.}}}

\subsection{RQ2: Why \sys Gains}
\label{sec:eval-time-analysis}

\begin{figure}[t]
  \centering
  \vspace{-1.5em}
  \includegraphics[width=0.94\columnwidth]{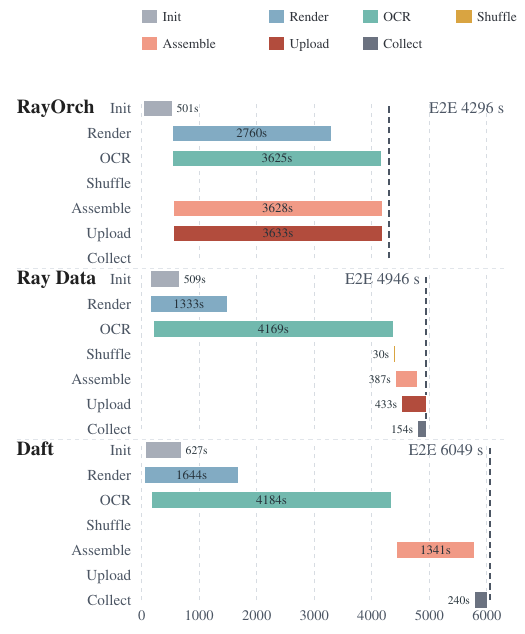}
  \vspace{-1.0em}
\caption{Cold end-to-end MinerU timelines on 64 NVIDIA H20 GPUs and
  174,744 valid pages.  Bars show concurrent stage windows; dashed lines mark
  E2E completion.  \sys overlaps assembly/upload with OCR, avoiding the
  post-OCR tails visible in Ray Data and Daft.}
  \Description{Three timelines compare cold MinerU executions.  \sys
  finishes in 4295.7 seconds, Ray Data in 4945.8 seconds, and Daft in 6048.5
  seconds.  \sys overlaps rendering, OCR, assembly, and upload.  Ray Data
  has a short post-OCR shuffle and assembly tail, while Daft has a much longer
  post-OCR assembly and collect tail.}
  \label{fig:mineru-cold-timeline}
  \vspace{-1.0em}
\end{figure}

\vspace{0.5em}
\noindent\textbf{\sys overlaps assembly and upload with OCR.}
RQ2 examines the elapsed MinerU stage windows in
Figure~\ref{fig:mineru-cold-timeline} from the same cold 64-GPU runs reported
in Table~\ref{tab:mineru-e2e}; each window spans a stage's first start to last
completion, and concurrent windows overlap rather than add.  \sys's
3,625-second OCR window runs concurrently with its 3,628-second assembly and
3,633-second upload windows, and the run finishes in 4,295.7 seconds.  Once a
parent's lineage is complete, parent-local commit releases that result to the
running upload stage without waiting for unrelated parents.  Ray Data and
Daft instead flatten children and globally regroup them before final assembly,
leaving post-OCR shuffle, assembly, and collection tails before their
4,945.8-second and 6,048.5-second E2E endpoints.  \sys's shorter terminal
\emph{Collect} window is not omitted work: parents emit incrementally, and all
work remains in the E2E measurement.

\vspace{0.5em}
\noindent
{\setlength{\fboxsep}{4pt}%
\setlength{\fboxrule}{0.5pt}%
\fcolorbox{blue!45}{blue!5}{%
  \parbox{\dimexpr\columnwidth-2\fboxsep-2\fboxrule\relax}{%
    \textbf{Answer to RQ2.} \sys avoids the baselines' post-OCR regrouping
    tail by completing parents locally and overlapping assembly and upload
    with OCR.}}}

\subsection{RQ3: Scheduling Ablations}
\label{sec:eval-scheduling-ablation}

\vspace{0.5em}
\noindent\textbf{FIFO further reduces wall time beyond rebatching.}
Table~\ref{tab:scheduling-ablation} holds lineage and commit fixed while
cumulatively adding 1:$M$ rebatching and FIFO scheduling.  The 818.0-second
native document-streaming row is an architectural reference because the full
system differs in both rebatching and queueing; the controlled comparison is
therefore $+$Rebatching versus $+$FIFO (full).  With the OCR UDF and batch
limit fixed, enabling FIFO lowers wall time from 634.1 to 579.3 seconds, an
absolute saving of 54.8 seconds and an 8.6\% reduction.  Equivalently, the
variant without FIFO is 9.5\% slower than the full system.

\input{tables/scheduling_ablation}

\vspace{0.5em}
\noindent
{\setlength{\fboxsep}{4pt}%
\setlength{\fboxrule}{0.5pt}%
\fcolorbox{blue!45}{blue!5}{%
  \parbox{\dimexpr\columnwidth-2\fboxsep-2\fboxrule\relax}{%
    \textbf{Answer to RQ3.} FIFO scheduling reduces wall time by an additional
    8.6\% beyond 1:$M$ rebatching.}}}

\subsection{RQ4: Lineage-Scoped Failure Containment}
\label{sec:eval-failure-containment}

\vspace{0.5em}
\noindent\textbf{\sys suppresses doomed siblings before UDF execution.}
RQ4 uses four H20 GPUs, four actors, a batch limit of 48, and a fixed 50-ms
page UDF.  We inject a failure into page 0 of each of the 99 largest parents,
which account for 5.25\% of documents but 51.9\% of pages.  For each runtime,
we execute three clean--poisoned run pairs (18 runs total).  Ray Data and Daft
pre-expand and partition pages outside the timed region, carry parent/error
columns, and filter poisoned-parent outputs only after regrouping; this
equalizes final outputs but does not enable runtime sibling suppression.  All
18 runs produce the expected healthy outputs.  As
Table~\ref{tab:failure-containment} shows, \sys prevents 6,241
\textsc{ready} siblings (26.5\% of poisoned-parent siblings and 13.7\%
overall) from entering the UDF, reducing non-trigger UDF calls from 45,408 to
39,167.  Across the three pairs, \sys reduces paired wall time by
14.65--15.24\% (14.93\% mean), whereas Ray Data changes by 0.16--0.22\%
(0.18\% mean) and Daft by $-0.16$--1.19\% (0.36\% mean).

\input{tables/failure_containment}

\vspace{0.5em}
\noindent
{\setlength{\fboxsep}{4pt}%
\setlength{\fboxrule}{0.5pt}%
\fcolorbox{blue!45}{blue!5}{%
  \parbox{\dimexpr\columnwidth-2\fboxsep-2\fboxrule\relax}{%
    \textbf{Answer to RQ4.} Runtime-owned lineage suppresses doomed sibling
    work while preserving all expected healthy outputs.}}}


%% file: tables/evaluation_setup.tex
\begin{table}[H]
\caption{Evaluation workloads, scales, and configurations.}
\vspace{-1em}
\label{tab:eval-setup}
\small
\centering
\renewcommand{\arraystretch}{1.12}
\setlength{\tabcolsep}{2.5pt}
\begin{tabular}{>{\raggedright\arraybackslash}m{0.15\columnwidth}|
                  >{\centering\arraybackslash}m{0.34\columnwidth}|
                  >{\centering\arraybackslash}m{0.38\columnwidth}}
\hline
\textbf{Workload} & \textbf{Input scale} & \textbf{Pipeline / hardware} \\ \hline
\textbf{MinerU} & 3,689 PDFs / 174,744 pages & render~$\to$~OCR~$\to$~assemble; 4--64 H20 GPUs \\ \hline
\textbf{Video} & 27,091 videos / 104,952 clips & decode~$\to$~model~$\to$~reduce; 8--64 H20 GPUs \\ \hline
\textbf{Docling} & 2,000 PDFs & document preparation; 4 H20 GPUs \\ \hline
\textbf{Ablation} & 368 PDFs / 7,072 pages & streaming~$\to$~rebatch~$\to$~FIFO; 4 H20 GPUs \\ \hline
\textbf{Failure trace} & 1,885 PDFs / 45,507 pages & fixed-cost page UDF; 4 H20 GPUs \\ \hline
\multicolumn{3}{l}{\scriptsize Ray 2.50.0; Daft 0.7.21; Torch 2.7.1; MinerU2.5-2509-1.2B; Qwen2.5-VL-7B.} \\ \hline
\end{tabular}
\vspace{-1em}
\end{table}

%% file: tables/mineru_e2e.tex
\begin{table}[t]
  \centering
  \caption{MinerU end-to-end performance on 64 NVIDIA H20 GPUs.  Lower E2E
    time and higher throughput are better.}
    \vspace{-1em}
  \label{tab:mineru-e2e}
  \small
  \setlength{\tabcolsep}{2.5pt}

  \begin{tabular}{l|r|r}
    \hline
    \textbf{System}
      & \multicolumn{1}{c|}{\textbf{E2E Time (s)} $\downarrow$}
      & \multicolumn{1}{c}{\textbf{Throughput (page/s)} $\uparrow$} \\ \hline
    \textbf{RayOrch}
      & \bfseries 4295.7
      & \bfseries 40.6788 \\ \hline
    \textbf{Ray Data}
      & 4945.8
      & 35.3318 \\ \hline
    \textbf{Daft}
      & 6048.5
      & 28.8903 \\ \hline
    \textbf{Native MinerU}
      & 8874.47
      & 19.6907 \\ \hline
  \end{tabular}

  \vspace{-1.0em}
\end{table}

%% file: tables/docling_e2e.tex
\begin{table}[t]
  \centering
  \caption{Docling end-to-end performance on four NVIDIA H20 GPUs.  Lower E2E
  time and higher throughput are better.}
  \vspace{-1.0em}
  \label{tab:docling-e2e}
  \small
  \setlength{\tabcolsep}{2.5pt}

  \begin{tabular}{l|r|r}
    \hline
    \textbf{System}
      & \multicolumn{1}{c|}{\textbf{E2E Time (s)} $\downarrow$}
      & \multicolumn{1}{c}{\textbf{Throughput (doc/s)} $\uparrow$} \\ \hline
    \textbf{RayOrch}
      & \bfseries 9489.07
      & \bfseries 0.2107 \\ \hline
    \textbf{Ray Data}
      & 11298.75
      & 0.1769 \\ \hline
    \textbf{Docling Serve}
      & 12214.70
      & 0.1637 \\ \hline
  \end{tabular}
  \vspace{-1.0em}
\end{table}

%% file: tables/scheduling_ablation.tex
\begin{table}[h]
\vspace{-1em}
\caption{Cumulative scheduler ablation; slowdown is relative to the full system.}
\label{tab:scheduling-ablation}
\small
\centering
\setlength{\tabcolsep}{2.0pt}
\renewcommand{\arraystretch}{1.05}
\begin{tabular}{l|c|c|c|r|r}
\hline
\textbf{Variant} & \textbf{Stream} & \textbf{1:$M$} & \textbf{FIFO} & \textbf{Wall (s)} & \textbf{Slowdown} \\ \hline
\textbf{Document streaming$^{*}$} & \ding{51} & $\times$ & $\times$ & 818.0 & +41.2\% \\ \hline
\textbf{$+$Rebatching} & \ding{51} & \ding{51} & $\times$ & 634.1 & +9.5\% \\ \hline
\textbf{$+$FIFO (full)} & \ding{51} & \ding{51} & \ding{51} & 579.3 & 0.0\% \\ \hline
\end{tabular}
\vspace{-1em}
\end{table}

%% file: tables/failure_containment.tex
\begin{table}[H]

\caption{Parent-scoped failure containment over three runs.}
\label{tab:failure-containment}
\vspace{-1.0em}
\small
\centering
\setlength{\tabcolsep}{1.6pt}
\renewcommand{\arraystretch}{1.0}
\begin{tabular}{l|c|r|r|r}
\hline
\textbf{System} & \shortstack{\textbf{Sibling}\\\textbf{suppression}} &
\shortstack{\textbf{Siblings}\\\textbf{not dispatched}} &
\shortstack{\textbf{Non-trigger}\\\textbf{UDF calls}} &
\shortstack{\textbf{Time}\\\textbf{reduction}} \\ \hline
\textbf{RayOrch} & \ding{51} & 6,241 (26.5\%) & 39,167 & 14.93\% \\ \hline
\textbf{Ray Data} & $\times$ & 0 & 45,408 & 0.18\% \\ \hline
\textbf{Daft} & $\times$ & 0 & 45,408 & 0.36\% \\ \hline
\end{tabular}
\vspace{-1.0em}
\end{table}

%% file: sections/8_conclusion.tex
\section{Conclusion}
\label{sec:conclusion}

Foundation-model data-preparation pipelines repeatedly change processing granularity while preserving parent-child membership, order, completion, and failure scope. \sys addresses this challenge with a programming model and Ray-based engine for finite, acyclic multi-grain dataflows. Compiler-validated expansions and parent-scoped gathers, runtime-owned structural lineage, per-Call FIFO queues, generation-fenced commits, and typed failure containment enable cross-parent batching, ordered local reconstruction, independent parent completion, and isolated recovery. On NVIDIA H20 GPUs, \sys reduced end-to-end time by 13.1\% versus Ray Data and 29.0\% versus Daft on MinerU, and by 16.0\% versus Ray Data on Docling, while achieving strong-scaling speedups of $15.14\times$ and $7.82\times$ through 64 GPUs on MinerU and video, respectively. FIFO scheduling further reduced wall time by 8.6\%, and lineage-scoped containment suppressed 6,241 non-trigger sibling computations while preserving all expected outputs for unaffected parents. 